# The Evolution of Overconfidence

Dominic D. P. Johnson[1] & James H. Fowler[2]

[1] *Politics & International Relations, University of Edinburgh, UK*

[2] *Division of Medical Genetics and Department of Political Science, University of California, San Diego, USA*

**Confidence is an essential ingredient of success in a wide range of domains ranging from job performance and mental health, to sports, business, and combat[1-4]. Some authors have suggested that not just confidence but *overconfidence*—believing you are better than you are in reality—is advantageous because it serves to increase ambition, morale, resolve, persistence, or the credibility of bluffing, generating a self-fulfilling prophecy in which exaggerated confidence actually increases the probability of success[3-8]. However, overconfidence also leads to faulty assessments, unrealistic expectations, and hazardous decisions, so it remains a puzzle how such a false belief could evolve or remain stable in a population of competing strategies that include accurate, unbiased beliefs. Here, we present an evolutionary model showing that, counter-intuitively, overconfidence maximizes individual fitness and populations will tend to become overconfident, as long as benefits from contested resources are sufficiently large compared to the cost of competition. In contrast, "rational" unbiased strategies are only stable under limited conditions. The fact that overconfident populations are evolutionarily stable in a wide range of environments may help to explain why overconfidence remains prevalent today, even if it contributes to hubris, market bubbles, financial collapses, policy failures, disasters, and costly wars[9-13].**

Humans exhibit many psychological biases, but one of the most consistent, powerful, and widespread is overconfidence. Most people show a bias towards: (1)

exaggerated personal qualities and capabilities; (2) an illusion of control over events; and (3) invulnerability to risk (three phenomena collectively known as "positive illusions")[2-4,14]. Overconfidence amounts to an "error" of judgment or decision-making, because it leads to *overestimating* one's capabilities, and/or *underestimating* an opponent, the difficulty of a task, or possible risks. It is therefore no surprise that overconfidence has been blamed throughout history for high-profile disasters such as World War I, the Vietnam war, the war in Iraq, the 2008 financial crisis, and the ill preparedness for environmental phenomena like Hurricane Katrina and climate change[9,12,13,15,16].

If overconfidence is both a widespread feature of human psychology *and* causes costly mistakes, we are faced with an evolutionary puzzle as to why humans should have evolved or maintained such an apparently damaging bias. One possible solution is that overconfidence can actually be advantageous on average (even if costly at times), because it boosts ambition, morale, resolve, persistence, or the credibility of bluffing. If such features increase net payoffs in competition or conflict, then overconfidence may have been favoured by natural selection over the course of human evolutionary history[5-8].

However, it is unclear whether such a bias can evolve in realistic competition with alternative strategies. The null hypothesis is that biases would die out, because they lead to faulty assessments and suboptimal behaviour. In fact, a large class of economic models depend on the assumption that biases in beliefs do not exist[17]. Underlying this assumption is the idea that there must be some evolutionary or learning process that causes individuals with *correct* beliefs to be rewarded (and thus to spread at the expense of incorrect beliefs). However, unbiased decisions are not necessarily the best strategy for maximizing benefits over costs, especially under conditions of competition, uncertainty, and asymmetric costs of different types of errors[8,18-21]. While economists tend to posit the notion of human brains as general-purpose utility maximizing machines that evaluate the costs, benefits, and probabilities of different options on a case-by-case basis, natural selection may have favoured the development of simple heuristic biases (such as overconfidence) in a given domain because they were more economical, available, or faster.

Here, we present a model showing that, under plausible conditions for the value of rewards, the cost of conflict, and uncertainty about the capability of competitors, there can be material rewards for holding *incorrect* beliefs about one's own capability. These adaptive advantages of overconfidence may explain its emergence and spread in humans, other animals or indeed any interacting entities (by a process of trial and error, imitation, learning, or selection). The scenario we model—a competition for resources—is simple but general, thereby capturing the essence of a broad range of competitive interactions including animal conflict, strategic decision-making, market competition, litigation, finance, and war.

Suppose a resource $r$ is available to an individual that claims it, and there are two individuals, $i$ and $j$. These individuals each have initial "capability" $\theta_i$ and $\theta_j$ that determine whether or not they would win a conflict over the resource. Without loss of generality, we assume $\theta$ is distributed in the population according to a symmetric stable probability density[22] with cumulative distribution $\Phi$, mean 0, and variance $\sqrt{1/2}$. The initial advantage to individual $i$ is $a = \theta_i - \theta_j$, and assumptions about the distribution of $\theta$ imply that the probability density of $a$ has a cumulative distribution $\Phi$, mean 0, and unit variance (Supplementary Information Sec. 1).

If neither individual claims the resource then no fitness is gained. If only one makes a claim, then the claimant acquires the resource and gains fitness $r$ and the other individual gains nothing. If both claim the resource, then both pay a cost $c$ due to the conflict between them, but the individual with the higher initial capability will win the conflict, acquiring the resource and obtaining fitness $r$. This means there are only three outcomes that have an impact on an individual's fitness: winning a conflict ($W$); losing a conflict ($L$); and obtaining an unclaimed resource ($O$). Given the probability of each of these outcomes ($p_W$, $p_L$ and $p_O$), the benefits of obtaining the resource $r$, and the costs of conflict $c$, the expected fitness is $E(f) = p_W(r-c) + p_L(-c) + p_O(r)$. Note that $r$ and $c$ can denote *expected* benefits and costs—if conflict outcomes were made probabilistic instead of deterministic, the results would not change.

Individuals choose whether or not to claim a resource based on their perceived capability relative to the capability of other claimants. If there were no uncertainty in this assessment, there would never be a conflict because the dispute can be settled without cost (the stronger individual takes the resource, and the weaker individual surrenders it, allowing both agents to avoid $c$)[23-26]. In the real world, however, uncertainty is common. We therefore model individuals' uncertainty about their opponent's capability by adding an error term $v$ to the opponent's capability such that individual $i$ thinks the capability of individual $j$ is $\theta_j + v_i$. To derive analytical results, we assume that this perception error has a magnitude of $\varepsilon > 0$ and is binomially distributed, with $\Pr(v = \varepsilon) = \Pr(v = -\varepsilon) = 0.5$ (the "binomial model"). To evaluate the role of confidence, we allow individuals to perceive their own capability as $\theta + k$, where $k = 0$ indicates *unbiased* individuals who perceive their capability correctly, $k > 0$ indicates *overconfident* individuals who think they are stronger than they actually are, and $k < 0$ indicates *underconfident* individuals who think they are weaker than they actually are.

We explore the emergence and stability of biases in hypothetical populations using standard assumptions about evolutionary dynamics[27] under which the most fit are more likely to survive or reproduce, or the less fit are likely to copy better strategies. Fig. 1a shows regions of the parameter space and five equilibria that occur in the binomial model, all confirmed both analytically and numerically (see Supplementary Information Sec. 2).

When $r/c > 3/2$, the unique equilibrium is a pure (monomorphic) population of overconfident individuals, all of whom evolve a level of overconfidence that is equal to the size of the perception error ($k^* = \varepsilon$). As long as there is at least some perception error, overconfident individuals resist invasion by all other individuals, including underconfident ($k < 0$), unbiased ($k = 0$), and other kinds of overconfident individuals ($k > 0$).

When $1/3 < r/c < 3/2$, there are two equilibria. First, a mixed (polymorphic) population made up of overconfident individuals ($k^* = \varepsilon$) and underconfident individuals ($k^* = -\varepsilon$) is always possible as long as there is at least some perception error. Second, an unbiased equilibrium ($k^* = 0$), is also possible in this region, but only if the perception error is sufficiently low.

Finally, when $r/c < 1/3$ there are two more equilibria. A pure equilibrium of underconfident individuals ($k^* = -\varepsilon$) is always possible, and a mixed equilibrium of very underconfident ($k^* = -2\varepsilon$) and unbiased ($k^* = 0$) individuals is possible when there is a moderate amount of uncertainty.

The underlying assumptions of the binomial model are deliberately simple in order to make closed-form characterizations tractable. We also used numerical simulation methods to evaluate the model when we allow the perception error $\nu$ to vary continuously using a normal distribution with mean 0 and standard deviation $\varepsilon$ (the "normal model"). This assumption may be more realistic than the binomial assumption since it allows perception errors to vary in magnitude.

As with the binomial model, the normal model shows that overconfidence ($k^* > 0$) is the unique pure equilibrium when the benefit/cost ratio is high enough (approximately $r/c > 0.7$, see Fig. 1b), which is notably less stringent than the binomial model reported above. When the benefit/cost ratio falls below this critical value, the unique pure equilibrium is underconfidence ($k^* < 0$). If there is any perception error whatsoever, an *absence* of bias is only an equilibrium at a single point—the value of resources and the cost of conflict must be in perfect balance to eliminate bias (Fig.1 b). This result suggests that models based on the assumption that individuals perceive their own capabilities without bias[17] are unrealistic—any small change in the benefit/cost ratio will tilt the advantage away from "rational" individuals towards those that assume they are more or less capable than they really are.

The normal model also yields the same positive relationship between perception error and confidence that was derived in the binomial model. As uncertainty about opponent capabilities increases, it becomes more advantageous to express stronger bias (the overconfident become even more confident and the underconfident become even less confident).

The simulations allowed us to examine some extensions of the model (see Supplementary Information Sec. 3). If we generalize the model to 3 players, overconfidence is favoured at the same threshold ($r/c > 0.7$). Results are also robust if we allow conflict costs to vary between winners and losers. The threshold required for overconfidence decreases as losers suffer relatively more. For example, overconfidence evolves when $r/c > 0.6$ if the costs to the winner are $0.8c$, and it evolves when $r/c > 0.45$ if the costs to the winner are $0.2c$. In other words, when conflict for the winner is cheap, overconfidence is even more likely to evolve and persist.

Our model shares interesting parallels with the famous Hawk-Dove game in evolutionary game theory[24]. "Hawks" escalate until they win (with benefit $b$), or sustain significant injury (with cost $c$). "Doves" only display, and retreat if attacked. Where $b > c$, Hawks take over the population and animals always fight. Where $c > b$, a mixed population of Hawks and Doves emerges. The Hawk-Dove game is important because it shows that (where $c > b$) contests can be resolved by "conventional" signals (displays only) with minimal fighting—explaining why many animals have dangerous weapons (such as sharp horns or teeth) but death is rare.

We find that the Hawk-Dove game is a special case of our model, where the only possible strategies are to be infinitely overconfident ($k = \infty$, i.e. Hawk) and therefore always claim the resource, or infinitely underconfident ($k = -\infty$, i.e. Dove) and therefore never claim. In Supplementary Information (Sec. 4) we show that the standard equilibria of the Hawk-Dove game emerge under these conditions. Strikingly, however, somewhat overconfident (but not infinitely overconfident) individuals always beat both Hawk and Dove. Our model, therefore, shows that individuals with a more nuanced strategy—even a biased one—do better than the "extreme" strategies of Hawk and Dove. Moreover, hawkish (overconfident) strategies can dominate even where $c > r$, a finding that contrasts with previous Hawk-Dove models.

One important implication of the model is that environments with more valuable resources will generate *more* conflict (Supplementary Information Sec. 5). This parallels the finding in the animal fighting literature that, where very valuable resources are at stake, hawkish strategies become more common and, in contrast to much animal conflict that is ritualized and restrained, fighting under these conditions can become lethal[28].

The analysis here demonstrates that overconfidence often prevails over accurate assessment. Overconfidence is advantageous because it encourages individuals to claim resources they could not otherwise win if it came to a conflict (stronger but cautious rivals will sometimes fail to make a claim) and it keeps them from walking away from resources they would surely win. These results conform with previous observations that systematic overestimates of the probability of winning simple gambling games can be adaptive if the benefits of the resource at stake sufficiently exceed the costs of attempting to gain it[19,20], that aggressive strategies (like "Hawk" in Hawk-Dove games) are favoured if the advantages of winning exceed the costs of injury[24], and that overconfident states outperform others in an agent-based model of conflict[29].

Note that overconfidence in our model is purely *self*-deception—there is no *other*-deception ("bluffing") because there is no signalling of *k* (opponents are not gullible to others' inflated beliefs). This is important because it demonstrates that there are adaptive advantages of overconfidence irrespective of any possible (additional) advantages of bluffing. Bluffing is often argued to be unstable in nature because there would be strong selection on discriminating responses. However, this may be partly why *self*-deception evolved, in Robert Trivers' words: "hiding the truth from yourself to hide it more deeply from others"[6,7]. Previous work has also shown that bluffing can survive counter-selection if there is ambiguity in one's own or others' strengths. If so, bluffs and reality cannot be reliably distinguished, and calling others' bluff takes on a cost of its own. Maynard Smith and Parker (1976) suggested that bluffing is therefore *more* likely (even if it is detectable in principle) among animals in which serious injury is possible—i.e. those with weapons—because the costs of calling a bluff can be high.

Our model applies to any replicating entity or any species, but it has particular implications for humans. First, if contested resources were sufficiently valuable compared to the costs of competing for them during human evolutionary history, we might expect humans to have evolved a bias towards overconfidence[5,12,19,20]. Such an outcome is exactly what the experimental psychology literature has long demonstrated, and yet has lacked an explanation for its origin[2-4,14]. A recent review of whether *any* "false beliefs" could be biologically adaptive concluded that there is just a single compelling candidate: positive illusions[8]. Today, we may retain evolved proximate mechanisms that give rise to overconfidence even in situations where the costs of conflict have increased relative to the value of the reward, making it maladaptive in many modern settings (such as, perhaps, in interpersonal aggression and war).

Second, overconfidence may arise and spread more quickly among humans than other organisms. Rather than relying on genetic mutation and natural selection over many generations, overconfidence in humans can emerge and spread much more rapidly by other means such as trial and error, imitation, or learning (which may also generate considerable variation among different "ecological" contexts such as habitats, cultures, or organizations). These processes of cultural selection may affect the way strategies emerge, survive and spread today among interacting entities, whether individuals, groups, negotiators, lawyers, traders, banks, sports teams, firms, armies, or states. In many of these settings, overconfidence may pay on average even though it only attracts attention when it causes costly disasters, or when the environment (the $r/c$ ratio) changes such that overconfidence begins to generate net costs.

Other recent models have explored the evolution of risk preferences[30] but in the present model individuals do not prefer or avoid risk—their heuristic is simply to assess capabilities and claim the resource if they perceive a capability gap. Interestingly, as we show in Supplementary Information (Sec. 6), this heuristic causes individuals to behave *as if* they were calculating the expected outcome of a risky choice under a specific set of assumptions about themselves and their opponents and comparing it to a required risk premium, which is cognitively a much more demanding task. Thus, although it is possible that risk preferences contribute to behaviour in competition and conflict, the simpler mechanism of overconfidence provides a short-cut that yields equivalent outcomes. Such short-cuts may have been favoured in our evolution because they have lower operating costs, were more easily available to natural selection, or reach decisions faster. In fact, there are many examples of biases in human judgment and decision-making that appear to be adaptive precisely because they offer simple heuristics that *deceive* us into fitness maximizing behaviour[18,20].

The finding that the optimal level of bias increases with the magnitude of uncertainty is especially intriguing. It suggests that we should expect extreme levels of overconfidence (hubris) or underconfidence (fear) precisely when we are dealing with unfamiliar or poorly understood strategic contexts. We predict that—where the value of a prize sufficiently exceeds the costs of competing—overconfidence will be particularly prevalent in some very important domains that have inherently high levels of uncertainty, including: international relations (where events are complex, distant, and involve foreign cultures and languages), rare or unpredictable phenomena (such as natural disasters and climate change), novel or complex technologies (such as the internet bubble and modern financial instruments), and new and untested leaders, allies, and enemies. Although overconfidence may have been adaptive in our past, and may still be adaptive in some settings today, it appears that we are likely to become overconfident in precisely the most dangerous of scenarios.

**Supplementary Information** accompanies the paper.

**Acknowledgements:** We thank D. Blumstein, L.-E. Cederman, D. Fessler, P. Gočev, M. Haselton, D. Nettle, J. Orbell, K. Panchanathan, R. Trivers, N. Weidmann and R. Wrangham for discussions and help leading to this paper, and an anonymous referee for an impressive series of detailed analyses of our models. We have no competing interests to declare.

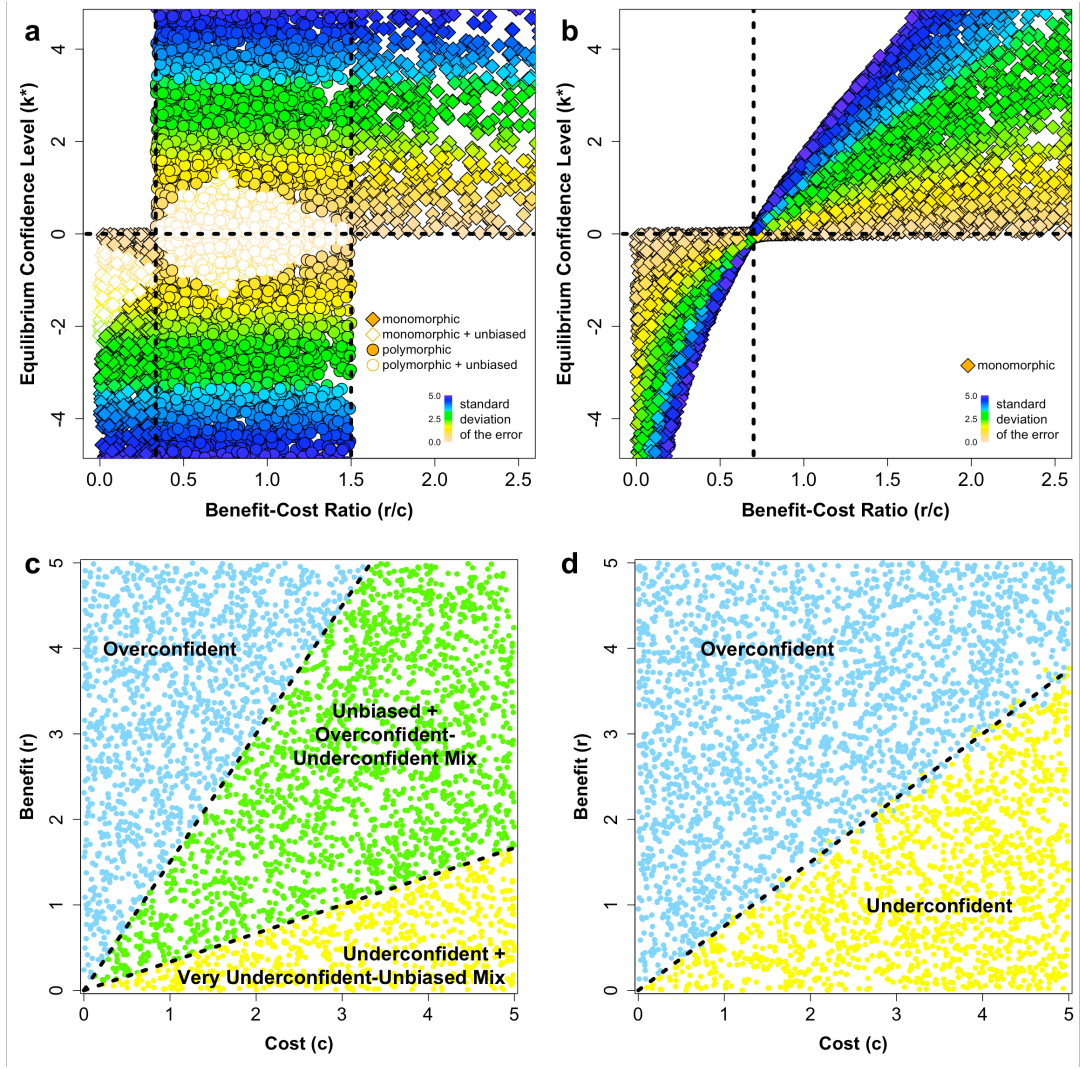

**Figure 1. Best performing levels of confidence across different parameter values.** The top panels show equilibrium levels of confidence $k^*$ for varying benefit/cost ratios (*r/c*) and degrees of uncertainty about the capabilities of competitors when assessment errors are modelled using a binomial distribution (**a**) or a normal distribution (**b**). Each point shows the results from a single simulation where the cost, benefit, and degree of uncertainty were drawn from a uniform distribution (see SI). There were 10,000 simulations in total. Shapes indicate types of equilibria that exist for a given parameter combination (filled shapes indicate unbiased strategies are not possible), and colours indicate the degree of uncertainty. The bottom panels show the same results for the binomial (**c**) and normal (**d**) models as a function of costs and benefits (colours indicate what kind of equilibria are possible; these results hold for all levels of perception error). Both models show that overconfident strategies are the unique equilibrium when the benefit/cost ratio is sufficiently high, and unbiased strategies are only possible under limited conditions.

# Supplementary Information for
# The Evolution of Overconfidence

Dominic D. P. Johnson

James H. Fowler

July 20, 2011

# Contents

# 1  The Model

Suppose a resource $r$ is potentially available to an individual that claims it. Suppose further that there are two individuals that can claim this resource, $i$ and $j$. These individuals each have a "capability" $\theta_i$ and $\theta_j$ which will determine whether or not they would win a conflict over the resource. Without loss of generality, we assume $\theta$ is distributed in the population according to a symmetric stable probability density[1] $\phi$ with cumulative distribution $\Phi$, mean 0, and variance $\sqrt{1/2}$. The endowment advantage to individual $i$ is $a = \theta_i - \theta_j$, and assumptions about the distribution of $\theta$ imply that the probability density of $a$ has a cumulative distribution of $\Phi$, mean 0, and variance 1.

If neither individual claims the resource then no fitness is gained. If only one individual makes a claim, then it acquires the resource and gains fitness $r$ and the other individual gains nothing. If both individuals claim the resource, then both individuals pay a cost $c$ due to the conflict between them, but the individual with the higher initial endowment will win the conflict and also obtain fitness $r$ for acquiring it.

An individual chooses whether or not to claim a resource based on only two pieces of information: their perception of their own capability and their perception of the capability of the other individual. If there were no uncertainty in this assessment, there would never be a conflict—the less capable individual would let the more capable individual claim the resource in order to avoid the costs of conflict in a fight they are bound to lose. We model an individual's uncertainty about their opponent's capability by adding a binomial error term: individual $i$ thinks the capability of individual $j$ is $\theta_j + \nu_i$. We assume that this perception error is unbiased, with $Pr(\nu = \epsilon) = 0.5$ and $Pr(\nu = -\epsilon) = 0.5$, and that perception errors of different individuals are independent of one another.

Assessments of capabilities can be wrong in two ways. First, there may be *errors* in the assessment of self or other. Second, there may be a *bias* in the assessment of self or other. Because it is relative capabilities that matter, any error or bias in assessments of oneself are equivalent to error or bias in assessments of others. So technically it does not matter to which individual each source of mistake is assigned. Here, however, we assign error to the assessment of others (because error is likely and realistic when one does not have full information about other agents), and bias to the assessment of self (because one does have full information about oneself, and yet psychologists have shown that individuals tend to be biased in self-assessments).

---

[1]This class of distributions includes the normal distribution. For more on these distributions, see Fama, E.F., Roll, R. 1968. Some properties of symmetric stable distributions. *Journal of the American Statistical Association* 63(323):817–836.

Notice that this setup is deliberately designed to capture the broadest and most generalizable elements of conflict of many types, including animal fighting, war, competition for resources, legal battles, bargaining, and so on. It allows us to model conflict over any resource of value, regardless of whether or not it is possessed – as long as a resource has not been consumed, it could potentially be "claimed" and derive benefit to the claimant. Notice also that although we assume that the outcome of conflict is a deterministic function of $a$, this is a random variable because of the capability distribution. Thus, we could equally assume that the probability of winning depends on a capability difference $b$ and a conflict-specific random variable $\psi$, and if this generates the same distribution ($Pr(a > 0) = Pr(b + \psi)$) then this probabilistic model will be equivalent to the deterministic model. Finally, notice that we do not assume conflict is a function of costs or benefits. As we will see, there *is* a relationship, but it arises endogenously as a consequence of other assumptions of the model.

To evaluate the role of confidence, we assume that individuals perceive their own capability as $\theta + k$, where $k = 0$ indicates *unbiased* individuals that perceive their capability correctly, $k > 0$ indicates *overconfident* individuals that think they are more capable than they actually are, and $k < 0$ indicates *underconfident* individuals that think they are less capable than they actually are.

There are, of course, many possible ways to model the mapping from capabilities and confidence to a decision under uncertainty (for example, one could assume that bias is a proportion of capability rather than a fixed amount that is independent of capability). One implication of the functional form we have chosen is that bias can apply equivalently to perception of one's own capability and perception of the opponent's capability (in other words, overconfidence can result either from under-estimating the opponent's capability or over-estimating one's own).

We assume individual $i$ only chooses to claim a resource if they think they could win a conflict over it ($\theta_i + k_i > \theta_j + \nu_i$), and conflict only occurs if the opponent also claims the resource (if $\theta_j + k_j > \theta_i + \nu_j$). Furthermore, when one individual claims a resource and the other does not, then the resource goes to the claimant. This means there are only three outcomes that have an impact on an individual's fitness, 1) winning a conflict ($W$), 2) losing a conflict ($L$), and 3) obtaining an unclaimed resource ($O$). The probabilities of these three events can be expressed as follows:

$$
\begin{aligned}
p_W =& Pr(\theta_i + k_i > \theta_j + \nu_i; \theta_j + k_j > \theta_i + \nu_j; \theta_i > \theta_j) = \\
& Pr(\nu_i - k_i < a; a < k_j - \nu_j; 0 < a) \\
p_L =& Pr(\theta_i + k_i > \theta_j + \nu_i; \theta_j + k_j > \theta_i + \nu_j; \theta_j > \theta_i) = \\
& Pr(\nu_i - k_i < a; a < k_j - \nu_j; a < 0) \\
p_O =& Pr(\theta_i + k_i > \theta_j + \nu_i; \theta_j + k_j < \theta_i + \nu_j) = \\
& Pr(\nu_i - k_i < a; a > k_j - \nu_j)
\end{aligned}
\tag{1}
$$

There are four equally likely conditions we need to check in order to specify an equation for each probability. For example:

$$
\begin{aligned}
p_W = \frac{1}{4}[ & p_W(\nu_i = \epsilon; \nu_j = \epsilon) + p_W(\nu_i = -\epsilon; \nu_j = \epsilon) + \\
& p_W(\nu_i = \epsilon; \nu_j = -\epsilon) + p_W(\nu_i = -\epsilon; \nu_j = -\epsilon)]
\end{aligned}
\tag{2}
$$

In some cases we also need to check whether the level of confidence $k$ is greater or less than the magnitude of the perception error $\epsilon$. For example, notice that when both individuals draw a positive $\epsilon$,

$$
\begin{aligned}
p_W(\nu_i = \nu_j = \epsilon) &= Pr(\epsilon - k_i < a; a < k_j - \epsilon; 0 < a) \\
&= Pr(\max(0, \epsilon - k_i) < a < k_j - \epsilon) \\
&= \begin{cases} \Phi(k_j - \epsilon) - \Phi(\epsilon - k_i) & \text{if } k_i \leq \epsilon \text{ and} \\ \Phi(k_j - \epsilon) - \frac{1}{2} & \text{otherwise.} \end{cases}
\end{aligned}
\tag{3}
$$

Thus it is a simple (but tedious) exercise to derive the complete probabilities by fixing $k$ relative to $\epsilon$ and then deriving the probability for each interval. For example, for monomorphic populations (where $k = k_i = k_j$) the probabilities reduce to:

$$p_W = p_L = \begin{cases} 0 & \text{if } k \leq -\epsilon \\ \frac{1}{4}\Phi(\epsilon + k) - \frac{1}{8} & \text{if } -\epsilon \leq k \leq 0 \\ \frac{1}{2}\Phi(\epsilon + k) - \frac{1}{4}(\epsilon - k) - \frac{1}{8} & \text{if } 0 \leq k \leq \epsilon \\ \frac{1}{2}\Phi(\epsilon + k) - \frac{1}{2}(\epsilon - k) & \text{if } \epsilon \leq k \end{cases}$$

$$p_O = \begin{cases} \frac{1}{2}\Phi(\epsilon + k) - \frac{1}{2}\Phi(\epsilon - k) + \frac{1}{2} & \text{if } k \leq -\epsilon \\ -\frac{1}{2}\Phi(\epsilon - k) + \frac{3}{4} & \text{if } -\epsilon \leq k \leq 0 \\ -\frac{1}{2}\Phi(\epsilon + k) + \frac{3}{4} & \text{if } 0 \leq k \leq \epsilon \\ -\frac{1}{2}\Phi(\epsilon + k) + \frac{1}{2}(\epsilon - k) + \frac{1}{2} & \text{if } \epsilon \leq k \end{cases} \qquad (4)$$

Given the benefits of obtaining the resource ($r$) and the costs of conflict ($c$), the expected fitness is

$$E(f) = p_W(r - c) + p_L(-c) + p_O(r) \qquad (5)$$

which for monomorphic populations simplifies to

$$= \begin{cases} \left(\frac{1}{2}\Phi(\epsilon + k) - \frac{1}{2}\Phi(\epsilon - k) + \frac{1}{2}\right)r & \text{if } k \leq -\epsilon \\ \left(\frac{1}{4}\Phi(\epsilon + k) - \frac{1}{2}\Phi(\epsilon - k) + \frac{5}{8}\right)r - \left(\frac{1}{2}\Phi(\epsilon + k) - \frac{1}{4}\right)c & \text{if } -\epsilon \leq k \leq 0 \\ \left(-\frac{1}{4}\Phi(\epsilon - k) + \frac{5}{8}\right)r - \left(\Phi(\epsilon + k) - \frac{1}{2}\Phi(\epsilon - k) - \frac{1}{4}\right)c & \text{if } 0 \leq k \leq \epsilon \\ \frac{1}{2}r - \left(\Phi(\epsilon + k) - \Phi(\epsilon - k)\right)c & \text{if } \epsilon \leq k \end{cases} \qquad (6)$$

In the next section we identify five equilibria that result from this model under the assumption that the perception error is binomial. We later use computer simulations to validate this analysis and then see what equilibria result when we change to an assumption that the perception error is normally distributed.

## 2   Five Equilibria of the Binomial Model

Here we present analytic characterizations of equilibria of the binomial model that resulted when we used numerical simulation to search the parameter space. To determine whether or not a population is in equilibrium, we checked to see whether an alternative strategy (an "invader") could survive in that population). This search revealed that fitness is always increasing in the confidence of the invader ($k_i$) for $k_i < -2\epsilon$ and decreasing for $k_i > 2\epsilon$. Graphical inspection of the fitness function shows that discontinuities appear at $k_i = \{-2\epsilon, -\epsilon, 0, \epsilon, 2\epsilon\}$ so we checked for equilibria for all pairwise combinations (including a strategy against itself) of these types. This search yielded three pure equilibria and two mixed equilibria as identified in the following sections.

### 2.1   Overconfident Populations ($k_j = \epsilon$)

Here we show that a population of individuals with $k_j = \epsilon$ resists invasion by *all possible types* ($-\infty < k_i < \infty$) as long as $\frac{r}{c} > \frac{3}{2}$.

We start with types that have a confidence level greater than the population overconfidence level ($k_i > \epsilon$). When an individual of type $k_i$ enters an overconfident population ($k_j = \epsilon$), the individual's expected fitness is:

$$\frac{1}{2}r - \left(\frac{1}{2}\Phi(2\epsilon) + \frac{1}{2}\Phi(k_i - \epsilon) + \frac{1}{2}\Phi(k_i + \epsilon) - \frac{3}{4}\right)c \tag{7}$$

Since $\Phi$ (the cumulative distribution of the capability difference) is increasing in $k_i$, higher values of $k_i$ yield greater expected costs, and invaders always do better when $k_i = \epsilon$ than when $k_i > \epsilon$. Therefore, no individual with $k_i > \epsilon$ can invade a population where $k_j = \epsilon$.

We can use the same reasoning to show that types with lower confidence $k_i < \epsilon$ are also incapable of invading. Given discontinuities in the fitness function, we need to check three conditions for a type $k_i$: $-\epsilon \leq k_i < \epsilon$, $-3\epsilon \leq k_i \leq -\epsilon$, and $k_i \leq -3\epsilon$.

When $-\epsilon \leq k_i < \epsilon$, the expected fitness of an invader is:

$$\left(\frac{1}{2}\Phi(k_i - \epsilon) + \frac{1}{4}\right)r - \left(\frac{1}{2}\Phi(2\epsilon) + \frac{1}{4}\Phi(k_i - \epsilon) + \frac{1}{2}\Phi(k_i + \epsilon) - \frac{5}{8}\right)c \tag{8}$$

This fitness is less than the average population fitness when:

$$\frac{r}{c} > \frac{\Phi(k_i + \epsilon) - \Phi(2\epsilon)}{\Phi(k_i - \epsilon) - \frac{1}{2}} + \frac{1}{2} \tag{9}$$

Note that the right hand side is less than $\frac{3}{2}$ if

$$\Phi(2\epsilon) - \Phi(k_i + \epsilon) + \Phi(k_i - \epsilon) - \frac{1}{2} < 0 \tag{10}$$

To test this condition for all $-\epsilon \leq k_i < \epsilon$, note there are three critical points, two at the boundaries ($k_i = -\epsilon$ and $k_i \to \epsilon$) and one where the first derivative equals zero:

$$\phi(k_i - \epsilon) - \phi(k_i + \epsilon) = 0 \tag{11}$$

Since $\phi$ is a symmetric stable distribution, this is only true when $|k_i - \epsilon| = |k_i + \epsilon|$, or $k_i = 0$.

An invader $k_i = -\epsilon$ cannot invade when $\frac{r}{c} > \frac{3}{2}$. To see why, we can use equations (1) and (5) to derive an expected fitness for this individual:

$$\left(\frac{3}{4} - \frac{1}{2}\Phi(2\epsilon)\right)r - \left(\frac{1}{4}\Phi(2\epsilon) - \frac{1}{8}\right)c \tag{12}$$

This fitness is always less than the average fitness for individuals in a population of overconfident individuals (equation (6)) if:

$$\frac{r}{c} > \frac{3}{2} \tag{13}$$

Similarly, for an invader $k_i = 0$ (unbiased) and we can use equations (1) and (5) to derive an expected fitness when such an individual enters a population of overconfident individuals ($k_j = \epsilon$):

$$\left(\frac{1}{2} - \frac{1}{2}\Phi(\epsilon)\right)r - \left(\frac{1}{4}\Phi(\epsilon) + \frac{1}{2}\Phi(2\epsilon) - \frac{3}{8}\right)c \tag{14}$$

This fitness is less than the average fitness for individuals in a monomorphic population of overconfident individuals when:

$$\frac{r}{c} > \frac{\Phi(2\epsilon) - \frac{1}{2}\Phi(\epsilon) - \frac{1}{4}}{\Phi(\epsilon)} \tag{15}$$

Notice that the numerator in this expression ranges with $\epsilon$ such that it never exceeds $\frac{1}{2}$ while the denominator is never less than $\frac{1}{2}$, meaning overconfident individuals are always safe if $\frac{r}{c} > 1$.

For invaders where $k_i \to \epsilon$, we must find the limit using L'Hôpital's rule. Notice that:

$$\lim_{k_i \to \epsilon} \frac{\Phi(k_i + \epsilon) - \Phi(2\epsilon)}{\Phi(k_i - \epsilon) - \frac{1}{2}} = \lim_{k_i \to \epsilon} \frac{\phi(k_i + \epsilon)}{\phi(k_i - \epsilon)} = \frac{\phi(2\epsilon)}{\phi(0)} \tag{16}$$

Since $\phi$ is symmetric with mean 0, the denominator lies at the highest point on the nonnegative function and the numerator lies below it if $\epsilon > 0$. Therefore, this value is less than one, and (9) implies that the condition $\frac{r}{c} > \frac{3}{2}$ is sufficient to ward off invasion for any type that approaches $k_i = \epsilon$.

Next, consider an invader $-3\epsilon \leq k_i \leq -\epsilon$. If this individual enters an overconfident population ($k_j = \epsilon$), the individual's expected fitness is:

$$\left(\frac{1}{2}\Phi(k_i + \epsilon) + \frac{1}{2}\Phi(k_i - \epsilon)\right) r - \left(\frac{1}{4}\Phi(2\epsilon) + \frac{1}{2}\Phi(k_i + \epsilon) - \frac{1}{4}\right) c \tag{17}$$

The fitness is increasing in $k_i$ if the first derivative is positive:

$$\left(\frac{1}{2}\phi(k_i + \epsilon) + \frac{1}{2}\phi(k_i - \epsilon)\right) r - \left(\frac{1}{2}\phi(k_i + \epsilon)\right) c > 0 \tag{18}$$

which is true if

$$\frac{r}{c} > \frac{1}{2}\left(\frac{\phi(k_i + \epsilon)}{\phi(k_i + \epsilon) + \frac{1}{2}\phi(k_i - \epsilon)}\right) \tag{19}$$

Since all terms are positive, $\phi(k_i + \epsilon)/(\phi(k_i + \epsilon) + \frac{1}{2}\phi(k_i - \epsilon)) < 1$ and the right hand side must be less than $\frac{1}{2}$. This means that if $\frac{r}{c} > \frac{1}{2}$ then fitness is maximized at $k_i = -\epsilon$, which we know from above cannot invade if $\frac{r}{c} > \frac{3}{2}$. Therefore, no type in the range with $-3\epsilon \leq k_i \leq -\epsilon$ can invade under the same conditions.

Finally, consider an invader with $k_i \leq -3\epsilon$. If this individual enters an overconfident population $(k_j = \epsilon)$, the individual's expected fitness is:

$$\left( \frac{1}{2}\Phi(k_i + \epsilon) + \frac{1}{2}\Phi(k_i - \epsilon) \right) r \tag{20}$$

Since $\Phi$ is increasing in $k_i$, fitness is maximized at $k_i = -3\epsilon$, which we know from above cannot invade if $\frac{r}{c} > \frac{3}{2}$. Therefore, no type $k_i \leq -3\epsilon$ can invade under the same conditions.

## 2.2   Underconfident Populations $(k_j = -\epsilon)$

Here we show that a population of individuals with $k_j = -\epsilon$ resists invasion by *all possible types* $(-\infty < k_i < \infty)$ as long as $\frac{r}{c} < \frac{1}{3}$.

We start with types that have a confidence level less than the population underconfidence level $(k_i < -\epsilon)$. The expected fitness for these individuals is:

$$\left( \frac{1}{2}\Phi(k_i + \epsilon) + \frac{1}{2}\Phi(k_i - \epsilon) \right) r \tag{21}$$

Since $\Phi$ is increasing in $k_i$, higher values of $k_i$ yield greater expected benefits, and invaders always do better when $k_i = -\epsilon$ than when $k_i < -\epsilon$. Therefore, no individual with $k_i < -\epsilon$ can invade a population where $k_j = -\epsilon$.

We can use the same reasoning to show that types with higher confidence $k_i > -\epsilon$ are also incapable of invading. To do so, we need to check three conditions: $-\epsilon < k_i \leq \epsilon$, $\epsilon \leq k_i \leq 3\epsilon$, and $k_i \geq 3\epsilon$.

When $-\epsilon < k - i \leq \epsilon$, the expected fitness of an invader that enters an underconfident population $(k_j = -\epsilon)$ is:

$$\left( \frac{1}{4}\Phi(k_i + \epsilon) + \frac{1}{2}\Phi(k_i - \epsilon) + \frac{1}{8} \right) r - \left( \frac{1}{4}\Phi(k_i + \epsilon) - \frac{1}{8} \right) c \tag{22}$$

This fitness is less than the average population fitness when:

$$\frac{c}{r} > 1 + 2\frac{\Phi(k_i - \epsilon) - \Phi(-2\epsilon)}{\Phi(k_i + \epsilon) - \frac{1}{2}} \tag{23}$$

Note that the right hand side is less than 3 if

$$\Phi(k_i - \epsilon) - \Phi(-2\epsilon) - \Phi(k_i + \epsilon) + \frac{1}{2} < 0 \tag{24}$$

To test this condition for all $-\epsilon < k_i \leq \epsilon$, note there are three critical points, two at the boundaries ($k_i = \epsilon$ and $k_i \to -\epsilon$) and one where the first derivative equals zero:

$$\phi(k_i - \epsilon) - \phi(k_i + \epsilon) = 0 \tag{25}$$

Since $\phi$ is a symmetric stable distribution, this is only true when $|k_i + \epsilon| = |k_i - \epsilon|$, or $k_i = 0$.

Substituting $k_i = \epsilon$ in the right hand side of equation (23) yields the condition $\frac{c}{r} > 3$, which means more confident types cannot invade unless:

$$\frac{r}{c} < \frac{1}{3} \tag{26}$$

Substituting $k_i = 0$ in the right hand side of equation (23) yields the condition:

$$\frac{c}{r} > 1 + 2\frac{\Phi(2\epsilon) - \Phi(\epsilon)}{\Phi(\epsilon) - \Phi(0)} \tag{27}$$

Notice that the right hand side here is less than 3 since the numerator is always greater than the denominator. Thus, the condition $\frac{r}{c} < \frac{1}{3}$ remains sufficient to ward off invaders.

For the point where $k_i \to -\epsilon$, we must find the limit using L'Hôpital's rule. Notice that:

$$\lim_{k_i \to -\epsilon} \frac{\Phi(k_i - \epsilon) - \Phi(-2\epsilon)}{\Phi(k_i + \epsilon) - \frac{1}{2}} = \lim_{k_i \to -\epsilon} \frac{\phi(k_i - \epsilon)}{\phi(k_i + \epsilon)} = \frac{\phi(-2\epsilon)}{\phi(0)} \tag{28}$$

Since $\phi$ is symmetric with mean 0, the denominator lies at the highest point on the nonnegative function and the numerator lies below it if $\epsilon < 0$. Therefore, this value is less than one, and (23) implies that the condition $\frac{r}{c} > \frac{1}{3}$ is sufficient to ward off invasion for any invader that approaches $k_i = -\epsilon$.

Next, consider an invader of type $\epsilon \leq k_i \leq 3\epsilon$. If this individual enters an underconfident population $(k_j = -\epsilon)$, the individual's expected fitness is:

$$\left(\frac{1}{4}\Phi(2\epsilon) + \frac{1}{4}\Phi(k_i - \epsilon) + \frac{1}{4}\right)r - \left(\frac{1}{2}\Phi(k_i + \epsilon) + \frac{1}{4}\Phi(k_i - \epsilon) - \frac{1}{4}\Phi(2\epsilon) - \frac{1}{2}\right)c$$

$$(29)$$

The fitness is decreasing in $k_i$ if the first derivative is negative:

$$\frac{1}{4}\phi(k_i - \epsilon)r - \left(\frac{1}{2}\phi(k_i + \epsilon) + \frac{1}{4}\phi(k_i - \epsilon)\right)c < 0 \tag{30}$$

which is true if

$$\frac{r}{c} < \left(\frac{2\phi(k_i + \epsilon) + \phi(k_i - \epsilon)}{\phi(k_i - \epsilon)}\right) \tag{31}$$

Since all terms are positive, the right hand side must be greater than 1. This means that if $\frac{r}{c} < 1$ then fitness is maximized at $k_i = \epsilon$, which we know from above cannot invade if $\frac{r}{c} < \frac{1}{3}$. Therefore, no invader with $\epsilon \leq k_i \leq 3\epsilon$ can invade under the same conditions.

Finally, consider an individual of type $k_i \geq 3\epsilon$. If this individual enters an underconfident population $(k_j = -\epsilon)$, the individual's expected fitness is:

$$\left(\frac{1}{2}\Phi(2\epsilon) + \frac{1}{4}\right)r - \left(\frac{1}{2}\Phi(k_i + \epsilon) + \frac{1}{4}\Phi(k_i - \epsilon) - \frac{1}{4}\Phi(2\epsilon) - \frac{1}{2}\right)c \tag{32}$$

Subtracting this from the underconfident $(k_i = -\epsilon)$ individual's fitness yields:

$$-\frac{1}{2}\left(\left(\Phi(k_i + \epsilon) - \frac{1}{2}\right) + \left(\Phi(k_i - \epsilon) - \frac{1}{2}\right) + \left(\Phi(2\epsilon) - \frac{1}{2}\right)\right)c \tag{33}$$

Which, by inspection, is always negative. Therefore, no type $k_i \geq 3\epsilon$ can invade under any conditions.

## 2.3   An Unbiased Population $(k_j = 0)$

Here we show the conditions under which a population of individuals with $k_j = 0$ resists invasion by *all possible types* $(-\infty < k_i < \infty)$. In particular the cost-benefit ratio must always fall within a range bounded by $\frac{1}{3} < \frac{r}{c} < \frac{3}{2}$, and this range gets smaller as $\epsilon$ increases, disappearing as soon as $\epsilon > 1.34$ if the capability distribution $\phi$ is normally distributed.

First, we use equations (1) and (5) to derive an expected fitness for an individual of type $k_i$ when entering a population of unbiased individuals:

$$
\begin{cases}
\left(\frac{1}{2}\Phi(k_i+\epsilon) + \frac{1}{2}\Phi(k_i-\epsilon)\right) r & \text{if } k_i \leq -2\epsilon \\[1.2em]
\begin{aligned} &\left(\frac{1}{2}\Phi(k_i+\epsilon) + \frac{1}{2}\Phi(k_i-\epsilon)\right) r \\ &- \left(\frac{1}{4}\Phi(k_i+\epsilon) + \frac{1}{4}\Phi(\epsilon) - \frac{1}{4}\right) c \end{aligned} & \text{if } -2\epsilon \leq k_i \leq -\epsilon \\[1.2em]
\begin{aligned} &\left(\frac{1}{4}\Phi(k_i+\epsilon) + \frac{1}{2}\Phi(k_i-\epsilon) + \frac{1}{8}\right) r \\ &- \left(\frac{1}{4}\Phi(k_i+\epsilon) + \frac{1}{4}\Phi(\epsilon) - \frac{1}{4}\right) c \end{aligned} & \text{if } -\epsilon \leq k_i \leq 0 \\[1.2em]
\begin{aligned} &\left(\frac{1}{2}\Phi(k_i-\epsilon) + \frac{1}{4}\Phi(\epsilon) + \frac{1}{8}\right) r \\ &- \left(\frac{1}{2}\Phi(k_i+\epsilon) + \frac{1}{4}\Phi(k_i-\epsilon) + \frac{1}{4}\Phi(\epsilon) - \frac{1}{2}\right) c \end{aligned} & \text{if } 0 \leq k_i \leq \epsilon \\[1.2em]
\begin{aligned} &\left(\frac{1}{4}\Phi(k_i-\epsilon) + \frac{1}{4}\Phi(\epsilon) + \frac{1}{4}\right) r \\ &- \left(\frac{1}{2}\Phi(k_i+\epsilon) + \frac{1}{4}\Phi(k_i-\epsilon) + \frac{1}{4}\Phi(\epsilon) - \frac{1}{2}\right) c \end{aligned} & \text{if } \epsilon \leq k_i \leq 2\epsilon \\[1.2em]
\begin{aligned} &\left(\frac{1}{2}\Phi(\epsilon) + \frac{1}{4}\right) r \\ &- \left(\frac{1}{2}\Phi(k_i+\epsilon) + \frac{1}{2}\Phi(k_i-\epsilon) - \frac{1}{2}\right) c \end{aligned} & \text{if } 2\epsilon \leq k_i
\end{cases}
\tag{34}
$$

Taking the derivative of this expression with respect to $k_i$ yields:

$$
\begin{cases}
\left(\frac{1}{2}\phi(k_i+\epsilon) + \frac{1}{2}\phi(k_i-\epsilon)\right) r & \text{if } k_i \leq -2\epsilon \\[1em]
\left(\frac{1}{2}\phi(k_i+\epsilon) + \frac{1}{2}\phi(k_i-\epsilon)\right) r - \frac{1}{4}\phi(k_i+\epsilon) c & \text{if } -2\epsilon \leq k_i \leq -\epsilon \\[1em]
\left(\frac{1}{4}\phi(k_i+\epsilon) + \frac{1}{2}\phi(k_i-\epsilon)\right) r - \frac{1}{4}\phi(k_i+\epsilon) c & \text{if } -\epsilon \leq k_i \leq 0 \\[1em]
\frac{1}{2}\phi(k_i-\epsilon) r - \left(\frac{1}{2}\phi(k_i+\epsilon) + \frac{1}{4}\phi(k_i-\epsilon)\right) c & \text{if } 0 \leq k_i \leq \epsilon \\[1em]
\frac{1}{4}\phi(k_i-\epsilon) r - \left(\frac{1}{2}\phi(k_i+\epsilon) + \frac{1}{4}\phi(k_i-\epsilon)\right) c & \text{if } \epsilon \leq k_i \leq 2\epsilon \\[1em]
- \left(\frac{1}{2}\phi(k_i+\epsilon) + \frac{1}{2}\phi(k_i-\epsilon)\right) c & \text{if } 2\epsilon \leq k_i
\end{cases}
\tag{35}
$$

We start with negative values of $k_i$. The previous equation shows that when $k_i \leq -2\epsilon$, fitness is always increasing in $k_i$. When $-2\epsilon \leq k_i \leq -\epsilon$ it is increasing as long as

$$\frac{r}{c} > \frac{1}{2 + 2\frac{\phi(k_i-\epsilon)}{\phi(k_i+\epsilon)}} \tag{36}$$

and therefore $k_i = -2\epsilon$ is a local maximum if

$$\frac{r}{c} < \frac{1}{2 + 2\frac{\phi(3\epsilon)}{\phi(\epsilon)}} \tag{37}$$

When $-\epsilon \leq k_i \leq 0$ fitness is increasing as long as

$$\frac{r}{c} > \frac{1}{1 + 2\frac{\phi(k_i-\epsilon)}{\phi(k_i+\epsilon)}} \tag{38}$$

and therefore $k_i = -\epsilon$ is a local maximum if

$$\frac{1}{2 + 2\frac{\phi(2\epsilon)}{\phi(0)}} < \frac{r}{c} < \frac{1}{1 + 2\frac{\phi(2\epsilon)}{\phi(0)}} \tag{39}$$

Finally, notice that since $\phi$ is a symmetric probability distribution with maximum value at $\phi(0)$, the term $\frac{\phi(k_i-\epsilon)}{\phi(k_i+\epsilon)}$ is increasing in $k_i$ when $-\epsilon \leq k_i \leq 0$. This means the most binding constraint in this range occurs at $k_i = 0$ where $\frac{\phi(k_i-\epsilon)}{\phi(k_i+\epsilon)} = 1$ and this point can only be a local maximum if $\frac{r}{c} > \frac{1}{3}$.

We now check positive values of $k_i$. For types $k_i \geq 2\epsilon$, fitness is always decreasing in $k_i$. For $\epsilon \leq k_i \leq 2\epsilon$ types it is decreasing as long as

$$\frac{r}{c} < 2\left(\frac{\phi(k_i+\epsilon)}{\phi(k_i-\epsilon)} - \frac{1}{2}\right) \tag{40}$$

and therefore $k_i = 2\epsilon$ is a local maximum if

$$\frac{r}{c} < \frac{2\phi(3\epsilon)}{\phi(\epsilon)} - 1 \tag{41}$$

For $0 \leq k_i \leq \epsilon$ types fitness is decreasing as long as

$$\frac{r}{c} < \frac{\phi(k_i+\epsilon)}{\phi(k_i-\epsilon)} - \frac{1}{2} \tag{42}$$

and therefore $k_i = \epsilon$ is a local maximum if

$$\frac{2\phi(3\epsilon)}{\phi(\epsilon)} - 1 < \frac{r}{c} < \frac{\phi(2\epsilon)}{\phi(0)} - \frac{1}{2} \tag{43}$$

Finally, notice that since $\phi$ is a symmetric probability distribution with maximum value at $\phi(0)$, the term $\frac{\phi(k_i+\epsilon)}{\phi(k_i-\epsilon)}$ is decreasing in $k_i$ when $0 \le k_i \le \epsilon$. This means the most binding constraint in this range occurs at $k_i = 0$ where $\frac{\phi(k_i+\epsilon)}{\phi(k_i-\epsilon)} = 1$ and this point can only be a local maximum if $\frac{r}{c} < \frac{3}{2}$.

To summarize, one requirement of equilibrium is that $k_i = 0$ be a local maximum, which can only be true if

$$\frac{1}{3} < \frac{r}{c} < \frac{3}{2} \tag{44}$$

Another requirement is that this local maximum be higher than the other local maxima at $k_i = \{-2\epsilon, -\epsilon, \epsilon, 2\epsilon\}$. To calculate these conditions, we can use the fitness equation in (34) to compare fitness of unbiased types vs. these other types. This yields the following conditions. Unbiased types do better than type $k_i = -2\epsilon$ if

$$\frac{r}{c} > \frac{1}{\frac{1}{2} + 2\left(\frac{\Phi(3\epsilon) - \Phi(0)}{\Phi(\epsilon) - \Phi(0)}\right)} \tag{45}$$

Unbiased types do better than type $k_i = -\epsilon$ if

$$\frac{r}{c} > \frac{1}{1 + 2\left(\frac{\Phi(2\epsilon) - \Phi(\epsilon)}{\Phi(\epsilon) - \Phi(0)}\right)} \tag{46}$$

Unbiased types do better than type $k_i = \epsilon$ if

$$\frac{r}{c} < \frac{\Phi(2\epsilon) - \Phi(\epsilon)}{\Phi(\epsilon) - \Phi(0)} + \frac{1}{2} \tag{47}$$

Unbiased types do better than type $k_i = 2\epsilon$ if

$$\frac{r}{c} < \frac{2}{3}\left(\frac{\Phi(3\epsilon) - \Phi(0)}{\Phi(\epsilon) - \Phi(0)}\right) \tag{48}$$

Notice that all of these constraints become more restrictive as $\epsilon$ increases. This is true because the derivative of the term $\frac{\Phi(3\epsilon)-\Phi(0)}{\Phi(\epsilon)-\Phi(0)}$ with respect to $\epsilon$ is

$$\frac{\left(\Phi(\epsilon)-\frac{1}{2}\right)\phi(3\epsilon)-\left(\Phi(3\epsilon)-\frac{1}{2}\right)\phi(\epsilon)}{\left(\Phi(\epsilon)-\frac{1}{2}\right)^2} \tag{49}$$

which is always negative since $\phi(\epsilon) > \phi(3\epsilon)$ and $\Phi(3\epsilon) > \Phi(\epsilon)$, and the derivative of the term $\frac{\Phi(2\epsilon)-\Phi(\epsilon)}{\Phi(\epsilon)-\Phi(0)}$ with respect to $\epsilon$ is

$$\frac{\left(\Phi(\epsilon)-\frac{1}{2}\right)(\phi(2\epsilon)-\phi(\epsilon))-(\Phi(2\epsilon)-\Phi(\epsilon))\phi(\epsilon)}{\left(\Phi(\epsilon)-\frac{1}{2}\right)^2} \tag{50}$$

which is always negative since $\phi(\epsilon) > \phi(2\epsilon)$. By inspection, as both of these terms become more negative, the constraints are more binding.

We calculated these constraints under the assumption that the symmetric stable distribution $\Phi$ is standard normal. Consistent with the numerical models below (see also Fig.1 in the main text), the results show that the highest $\epsilon$ that permits an unbiased population to ward off invasion is $\epsilon \approx 1.34$. This occurs at a cost-benefit ratio of $\frac{r}{c} \approx 0.71$ and at higher values of $\epsilon$ another type can always invade.

In other words, an unbiased population is only stable when the perception error $\epsilon$ is sufficiently low, *regardless of the costs of conflict or the benefits of the resource.*

## 2.4 Populations with a Mix of Overconfident Types ($k_j = \epsilon$) and Underconfident Types ($k_j = -\epsilon$)

Here we show the conditions under which a population with a mix of overconfident ($k_j = \epsilon$) and underconfident ($k_j = -\epsilon$) individuals resists invasion by *all possible types* ($-\infty < k_i < \infty$) as long as $\frac{1}{3} < \frac{r}{c} < \frac{3}{2}$.

At equilibrium, each type must obtain the same fitness:

$$pE(f|k_i = \epsilon, k_j = \epsilon) + (1-p)E(f|k_i = \epsilon, k_j = -\epsilon) =$$
$$pE(f|k_i = -\epsilon, k_j = \epsilon) + (1-p)E(f|k_i = -\epsilon, k_j = -\epsilon) \tag{51}$$

We use equations (1) and (5) to derive the payoffs for each type against itself and the opposing type, and use these to determine the equilibrium proportion $0 < p < 1$ of overconfident individuals and proportion $1 - p$ of underconfident individuals in the population. Setting the expected payoffs equal to one another yields:

$$\left(\frac{1}{2}r + \left(-\Phi(2\epsilon) + \frac{1}{2}\right)c\right)p+$$
$$\left(\left(\frac{1}{4}\Phi(2\epsilon) + \frac{3}{8}\right)r + \left(-\frac{1}{4}\Phi(2\epsilon) + \frac{1}{8}\right)c\right)(1-p) =$$
$$\left(\left(-\frac{1}{2}\Phi(2\epsilon) + \frac{3}{4}\right)r + \left(-\frac{1}{4}\Phi(2\epsilon) + \frac{1}{8}\right)c\right)p+\left(-\frac{1}{2}\Phi(2\epsilon) + \frac{3}{4}\right)r(1-p) \tag{52}$$

Simplifying and solving for $p$, we find the equilibrium mixture is

$$p = \frac{3r - c}{r + 2c} \tag{53}$$

and since $0 < p < 1$, it must also be true that

$$\frac{1}{3} < \frac{r}{c} < \frac{3}{2} \tag{54}$$

Note that $p$ is not conditional on the invading type $k_i$, which will be helpful in some cases in determining how fitness changes with respect to $k_i$.

For a population of types $-\epsilon$ and $\epsilon$, the fitness of an invader is the fitness against each type multiplied by the fraction of each type in the population: $pE(f|k_j = \epsilon) + (1-p)E(f|k_j = -\epsilon)$. Using equations (1) and (5) to derive the payoff of the invader against each type, we can see that when $k_i \geq 3\epsilon$,

the invader's fitness is

$$\left( \frac{1}{2}r + \left( -\frac{1}{2}\Phi(k_i + \epsilon) - \frac{1}{2}\Phi(k_i - \epsilon) - \frac{1}{2}\Phi(2\epsilon) + \frac{3}{4} \right) c \right) p +$$
$$\left( \left( \frac{1}{2}\Phi(2\epsilon) + \frac{1}{4} \right) r + \right.$$
$$\left. \left( -\frac{1}{2}\Phi(k_i + \epsilon) - \frac{1}{2}\Phi(k_i - \epsilon) + \frac{1}{2}\Phi(2\epsilon) + \frac{1}{4} \right) c \right)(1-p) \quad (55)$$

This equation is always decreasing in $k_i$ since every term is negative in which $k_i$ is involved as a positive element of the cumulative function $\Phi$.

When $\epsilon \leq k_i \leq 3\epsilon$, the payoff to the invader is

$$\left( \frac{1}{2}r + \left( -\frac{1}{2}\Phi(k_i + \epsilon) - \frac{1}{2}\Phi(k_i - \epsilon) - \frac{1}{2}\Phi(2\epsilon) + \frac{3}{4} \right) c \right) p +$$
$$\left( \left( \frac{1}{4}\Phi(2\epsilon) + \frac{1}{4}\Phi(k_i - \epsilon) + \frac{1}{4} \right) r + \right.$$
$$\left. \left( -\frac{1}{2}\Phi(k_i + \epsilon) - \frac{1}{4}\Phi(k_i - \epsilon) + \frac{1}{4}\Phi(2\epsilon) + \frac{1}{4} \right) c \right)(1-p). \quad (56)$$

Taking the first derivative with respect to $k_i$ and simplifying, we see that this payoff is decreasing when

$$\phi(k_i + \epsilon)\left( r(1-p) - 2c \right) - \phi(k_i - \epsilon)c(1-p) < 0 \quad (57)$$

which is always true if $r < 2c$ (or $\frac{r}{c} < 2$). Under these conditions, no $k_i > \epsilon$ can invade since the payoff is always decreasing as $k_i$ increases.

When $k_i \leq -3\epsilon$, the payoff to the invader is

$$\frac{1}{2}\left( \Phi(k_i + \epsilon) + \Phi(k_i - \epsilon) \right) r \quad (58)$$

which is always increasing in $k_i$.

Meanwhile, when $-3\epsilon \leq k_i \leq -\epsilon$, the payoff to the invader is

$$\frac{1}{2}\left( \Phi(k_i + \epsilon) + \Phi(k_i - \epsilon) \right) r - \frac{1}{4}\left( \Phi(2\epsilon) + \Phi(k_i + \epsilon) - 1 \right)cp \quad (59)$$

This payoff is increasing in $k_i$ if the first derivative of this equation is positive:

$$\frac{1}{2}\left(\phi(k_i+\epsilon)+\phi(k_i-\epsilon)\right)r - \frac{1}{4}\left(\phi(k_i+\epsilon)\right)cp > 0 \tag{60}$$

and substituting the equilibrium value of $p$ from equation (53), we get:

$$\phi(k_i+\epsilon)(2r^2+cr+c^2)+\phi(k_i-\epsilon)(3r-c) > 0 \tag{61}$$

Notice that this expression is always positive as long as $3r-c > 0$ (or $\frac{r}{c} > \frac{1}{3}$).

Finally, we check the interior to see what conditions must be met to prevent invaders of intermediate types from invading. The payoff to a type $-\epsilon \leq k_i \leq \epsilon$ is

$$\left(\left(\frac{1}{2}\Phi(k_i-\epsilon)+\frac{1}{4}\right)r + \left(-\frac{1}{2}\Phi(k_i+\epsilon)-\frac{1}{4}\Phi(k_i-\epsilon)-\frac{1}{2}\Phi(2\epsilon)+\frac{5}{8}\right)c\right)p+$$
$$\left(\left(\frac{1}{4}\Phi(k_i+\epsilon)+\frac{1}{2}\Phi(k_i-\epsilon)+\frac{1}{8}\right)r + \left(-\frac{1}{4}\Phi(k_i+\epsilon)+\frac{1}{8}\right)c\right)(1-p) \tag{62}$$

The first derivative of this expression is

$$\left(\left(\frac{1}{2}\phi(k_i-\epsilon)\right)r + \left(-\frac{1}{2}\phi(k_i+\epsilon)-\frac{1}{4}\phi(k_i-\epsilon)\right)c\right)p+$$
$$\left(\left(\frac{1}{4}\phi(k_i+\epsilon)+\frac{1}{2}\phi(k_i-\epsilon)\right)r + \left(-\frac{1}{4}\phi(k_i+\epsilon)\right)c\right)(1-p) \tag{63}$$

Substituting in the equilibrium value of $p$ from equation (53), this expression simplifies to

$$\frac{2r^2+cr+c^2}{4r+8c}\left(\phi(k_i-\epsilon)-\phi(k_i+\epsilon)\right) \tag{64}$$

Since $\phi$ is a symmetric probability distribution with maximum value at $\phi(0)$, the term $\phi(k_i-\epsilon)-\phi(k_i+\epsilon)$ is negative when $-\epsilon \leq k_i < 0$, 0 when $k_i = 0$, and positive when $0 < k_i \leq \epsilon$. This means that the maximum payoffs in this range are achieved at $k_i = -\epsilon$ and $k_i = \epsilon$ and therefore no interior type (including unbiased types) can invade.

## 2.5 Populations with a Mix of Very Underconfident Types ($k_j = -2\epsilon$) and Unbiased Types ($k_j = 0$)

Here we show the conditions under which a population with a mix of unbiased ($k_j = 0$) and very underconfident ($k_j = -2\epsilon$) individuals resists invasion by *all possible types* ($-\infty < k_i < \infty$) if $\epsilon$ is sufficiently small and $\frac{r}{c} < \frac{1}{3}$.

At equilibrium, each type must obtain the same fitness:

$$pE(f|k_i = -2\epsilon, k_j = 0) + (1-p)E(f|k_i = -2\epsilon, k_j = -2\epsilon) =$$
$$pE(f|k_i = 0, k_j = 0) + (1-p)E(f|k_i = 0, k_j = -2\epsilon) \quad (65)$$

We use equations (1) and (5) to derive the payoffs for each type against itself and the opposing type, and use these to determine the equilibrium proportion $0 < p < 1$ of overconfident individuals and proportion $1 - p$ of underconfident individuals in the population. Setting the expected payoffs equal to one another yields:

$$\left(1 - \frac{1}{2}\Phi(3\epsilon) + \frac{1}{2}\Phi(\epsilon)\right)rp + \left(1 - \frac{1}{2}\Phi(3\epsilon) + \frac{1}{2}\Phi(\epsilon)\right)r(1-p) =$$
$$\left(\left(-\frac{1}{4}\Phi(\epsilon) + \frac{5}{8}\right)r - \left(\frac{1}{2}\Phi(\epsilon) - \frac{1}{4}\right)c\right)p + \frac{1}{2}r(1-p) \quad (66)$$

Simplifying and solving for $p$, we find the equilibrium mixture is

$$p = \frac{2r}{r + 2c}\left(1 + \frac{\Phi(3\epsilon) - \frac{1}{2}}{\Phi(\epsilon) - \frac{1}{2}}\right) \quad (67)$$

Notice that $p$ is always positive as long as $\epsilon > 0$. And since $0 < p < 1$, it must also be true that for this equilibrium to exist

$$\frac{r}{c} < \frac{2}{1 + 2\frac{\Phi(3\epsilon) - \frac{1}{2}}{\Phi(\epsilon) - \frac{1}{2}}} \quad (68)$$

When $\epsilon \to \infty$, the c.d.f. $\Phi \to 1$, and this constraint simplifies to $\frac{r}{c} < \frac{2}{3}$. As $\epsilon$ decreases, the constraint becomes more binding. To determine the value when $\epsilon \to 0$, we must find the limit using L'Hôpital's rule:

$$\lim_{\epsilon \to 0} \frac{\Phi(3\epsilon) - \frac{1}{2}}{\Phi(\epsilon) - \frac{1}{2}} = \lim_{\epsilon \to 0} \frac{3\phi(3\epsilon)}{\phi(\epsilon)} = 3 \tag{69}$$

and the constraint for $p < 1$ simplifies to $\frac{r}{c} < \frac{2}{7}$.

For a population of types $0$ and $-2\epsilon$, the fitness of an invader is the fitness against each type multiplied by the fraction of each type in the population: $pE(f|k_j = 0) + (1-p)E(f|k_j = -2\epsilon)$. Using equations (1) and (5) to derive the payoff of the invader against each type, we can see that when $k_i \geq 4\epsilon$, the invader's fitness is

$$\left( \left( \frac{1}{2}\Phi(2\epsilon) + \frac{1}{4} \right) r + \left( -\frac{1}{2}\Phi(k_i + \epsilon) - \frac{1}{2}\Phi(k_i - \epsilon) + \frac{1}{2} \right) c \right) p$$
$$\left( \left( \frac{1}{2}\Phi(3\epsilon) + \frac{1}{2}\Phi(\epsilon) \right) r + \right.$$
$$\left. \left( -\frac{1}{2}\Phi(k_i + \epsilon) - \frac{1}{2}\Phi(k_i - \epsilon) + \frac{1}{2}\Phi(3\epsilon) + \frac{1}{2}\Phi(\epsilon) \right) c \right) (1-p). \tag{70}$$

This equation is always decreasing in $k_i$ since every term is negative in which $k_i$ is involved as a positive element of the cumulative function $\Phi$.

When $2\epsilon \leq k_i \leq 4\epsilon$, the payoff to the invader is

$$\left( \left( \frac{1}{2}\Phi(\epsilon) + \frac{1}{4} \right) r + \left( -\frac{1}{2}\Phi(k_i + \epsilon) - \frac{1}{2}\Phi(k_i - \epsilon) + \frac{1}{2} \right) c \right) p +$$
$$\left( \left( \frac{1}{4}\Phi(k_i - \epsilon) + \frac{1}{4}\Phi(3\epsilon) + \frac{1}{2}\Phi(\epsilon) \right) r + \right.$$
$$\left. \left( -\frac{1}{2}\Phi(k_i + \epsilon) - \frac{1}{4}\Phi(k_i - \epsilon) + \frac{1}{4}\Phi(3\epsilon) + \frac{1}{2}\Phi(\epsilon) \right) c \right) (1-p) \tag{71}$$

Taking the first derivative with respect to $k_i$ and simplifying, we see that this payoff is always decreasing because all terms are negative:

$$-3r^2 - \left( 2\frac{\Phi(3\epsilon) - \frac{1}{2}}{\Phi(\epsilon) - \frac{1}{2}} + 1 \right) rc - 2c^2 - \frac{2\phi(k_i + \epsilon)}{\phi(k_i - \epsilon)}(r + c) < 0 \tag{72}$$

When $\epsilon \leq k_i \leq 2\epsilon$, the payoff to the invader is

$$
\begin{aligned}
&\left(\left(\frac{1}{4}\Phi(k_i-\epsilon)+\frac{1}{4}\Phi(\epsilon)+\frac{1}{4}\right)r+\right.\\
&\left(-\frac{1}{2}\Phi(k_i+\epsilon)-\frac{1}{4}\Phi(k_i-\epsilon)-\frac{1}{4}\Phi(\epsilon)+\frac{1}{2}\right)c\bigg)p+\\
&\left(\left(\frac{1}{4}\Phi(k_i+\epsilon)+\frac{1}{2}\Phi(k_i-\epsilon)+\frac{1}{4}\Phi(\epsilon)\right)r+\right.\\
&\left.\left(-\frac{1}{4}\Phi(k_i+\epsilon)+\frac{1}{4}\Phi(\epsilon)\right)c\right)(1-p)
\end{aligned}
\tag{73}
$$

Taking the first derivative with respect to $k_i$ and simplifying, we see that this payoff is always decreasing because both terms are negative:

$$
-\frac{2(\Phi(3\epsilon)-\frac{1}{2})\left(\phi(k_i+\epsilon)+\phi(k_i-\epsilon)\right)}{\Phi(\epsilon)-\frac{1}{2}}r(r+c)-\Phi(k_i+\epsilon)(r^2+rc+2c^2)<0
\tag{74}
$$

When $0 \leq k_i \leq \epsilon$, the payoff to the invader is

$$
\begin{aligned}
&\left(\left(\frac{1}{2}\Phi(k_i-\epsilon)+\frac{1}{4}\Phi(\epsilon)+\frac{1}{8}\right)r+\right.\\
&\left(-\frac{1}{2}\Phi(k_i+\epsilon)-\frac{1}{4}\Phi(k_i-\epsilon)-\frac{1}{4}\Phi(\epsilon)+\frac{1}{2}\right)c\bigg)p+\\
&\left(\left(\frac{1}{4}\Phi(k_i+\epsilon)+\frac{1}{2}\Phi(k_i-\epsilon)+\frac{1}{4}\Phi(\epsilon)\right)r+\right.\\
&\left.\left(-\frac{1}{4}\Phi(k_i+\epsilon)+\frac{1}{4}\Phi(\epsilon)\right)c\right)(1-p)
\end{aligned}
\tag{75}
$$

Taking the first derivative with respect to $k_i$ and simplifying, we see that this payoff is decreasing when

$$
\begin{aligned}
&2\left(\phi(k_i+\epsilon)+\phi(k_i-\epsilon)\right)\left(r-\frac{\Phi(3\epsilon)-\frac{1}{2}}{\Phi(\epsilon)-\frac{1}{2}}c\right)r-\\
&\Phi(k_i+\epsilon)\left(3+2\frac{\Phi(3\epsilon)-\frac{1}{2}}{\Phi(\epsilon)-\frac{1}{2}}\right)r^2+rc+c^2<0
\end{aligned}
\tag{76}
$$

Notice that the second term is always negative, while the first term is negative if $r < \frac{\Phi(3\epsilon) - \frac{1}{2}}{\Phi(\epsilon) - \frac{1}{2}} c$. Since $\frac{\Phi(3\epsilon) - \frac{1}{2}}{\Phi(\epsilon) - \frac{1}{2}}$ ranges from $3 \to 1$ as $\epsilon$ goes from $0$ to $\infty$, this means that fitness is always decreasing in $k_i$ if $\frac{r}{c} < 1$.

Therefore, under the stricter constraint that $\frac{r}{c} < \frac{2}{3}$ to ensure that $p < 1$ for at least some values of $\epsilon$, fitness is decreasing for all invaders with positive $k_i$.

We now turn to invaders with values of $k_i \leq -2\epsilon$. Their payoff is

$$\frac{1}{2} \left( \Phi(k_i + \epsilon) + \Phi(k_i - \epsilon) \right) rp + \frac{1}{2} \left( \Phi(k_i + \epsilon) + \Phi(k_i - \epsilon) \right) r(1-p) \quad (77)$$

which is always increasing in $k_i$.

Finally, we check the interior between $k_i = -2\epsilon$ and $k_i = 0$ to see what conditions must be met to prevent invaders of intermediate types from invading. When $-2\epsilon \leq k_i \leq -\epsilon$, the payoff to the invader is

$$\left( \frac{1}{2} \left( \Phi(k_i + \epsilon) + \Phi(k_i - \epsilon) \right) r + \left( -\frac{1}{4} \Phi(k_i + \epsilon) - \frac{1}{4} \Phi(\epsilon) + \frac{1}{4} \right) c \right) p +$$
$$\frac{1}{2} \left( \Phi(k_i + \epsilon) + \Phi(k_i - \epsilon) \right) r(1-p) \quad (78)$$

If we subtract this payoff from the population payoff and rearrange, we find that this value is positive (invaders can invade) only when

$$\frac{r}{c} < \left( \frac{1}{1 + \frac{\Phi(k_i + \epsilon) - \Phi(-\epsilon)}{\Phi(k_i + \epsilon) - \Phi(-3\epsilon)}} \right) \left( \frac{\Phi(3\epsilon) - \frac{1}{2}}{\Phi(\epsilon) - \frac{1}{2}} \right) - 2 \quad (79)$$

Notice that in this range, $\Phi(k_i + \epsilon) - \Phi(-\epsilon) > \Phi(k_i - \epsilon) - \phi(-3\epsilon)$ if $\epsilon > 0$, and the difference in these differences increases in $k_i$ since $\Phi$ is convex in this range. Therefore, the constraint is decreasing in $k_i$ and becomes most binding when $k_i = -\epsilon$. The payoff to a type $-\epsilon \leq k_i \leq 0$ is

$$\left( \left( \frac{1}{4} \Phi(k_i + \epsilon) + \frac{1}{2} \Phi(k_i - \epsilon) + \frac{1}{8} \right) r + \left( -\frac{1}{4} \Phi(k_i + \epsilon) - \frac{1}{4} \Phi(\epsilon) + \frac{1}{4} \right) c \right) p +$$
$$\frac{1}{2} \left( \Phi(k_i + \epsilon) + \Phi(k_i - \epsilon) \right) r(1-p) \quad (80)$$

If we subtract this payoff from the population payoff and rearrange, we find that this value is positive (invaders can invade) only when

$$\frac{r}{c} < \frac{1}{\frac{1}{1+\frac{\Phi(-\epsilon)-\Phi(k_i-\epsilon)}{\Phi(\epsilon)-\Phi(k_i+\epsilon)}}\left(\frac{\Phi(3\epsilon)-\frac{1}{2}}{\Phi(\epsilon)-\frac{1}{2}}\right)-1} - 1 \tag{81}$$

Notice that in this range, $\Phi(-\epsilon) - \Phi(k_i - \epsilon)$ is increasing more quickly as $k_i$ increases because this difference lies on the concave part of $\Phi$, while $\Phi(\epsilon) - \Phi(k_i + \epsilon)$ is increasing more slowly because this difference lies on the convex part of $\Phi$. Therefore the constraint is increasing in $k_i$ and, as above, the constraint becomes most binding when $k_i = -\epsilon$.

For the last step, we check the conditions under which the population can resist invasion by a type $k_i = -\epsilon$. The fitness of this type against the population is

$$\left(\left(\frac{3}{4} - \frac{1}{2}\Phi(2\epsilon)\right)r + \left(\frac{1}{8} - \frac{1}{4}\Phi(\epsilon)\right)c\right)p + \left(\frac{3}{4} - \frac{1}{2}\Phi(2\epsilon)\right)r(1-p) \tag{82}$$

Subtracting this payoff from the population payoff yields an equation which must be positive to sustain equilibrium. Rearranging this equation yields the following constraint:

$$\frac{r}{c} < \frac{1}{1 - \frac{\Phi(2\epsilon)-\frac{1}{2}}{\Phi(3\epsilon)+\Phi(\epsilon)-1}} - 2 \tag{83}$$

Notice that this constraint approaches $\frac{r}{c} < 0$ as $\epsilon \to 0$ since

$$\lim_{\epsilon \to 0} \frac{\Phi(2\epsilon)-\frac{1}{2}}{\Phi(3\epsilon)+\Phi(\epsilon)-1} = \lim_{\epsilon \to 0} \frac{2\phi(2\epsilon)}{3\phi(3\epsilon)+\phi(\epsilon)} = \frac{2\phi(0)}{3\phi(0)+\phi(0)} = \frac{1}{2} \tag{84}$$

Similarly, the constraint approaches $\frac{r}{c} < 0$ as $\epsilon \to \infty$ since $\Phi(\infty) = 1$. For finite positive values, we can see that the constraint is positive. This requires $\frac{\Phi(2\epsilon)-\frac{1}{2}}{\Phi(3\epsilon)+\Phi(\epsilon)-1} > \frac{1}{2}$ which can also be written as $\Phi(2\epsilon) > \frac{\Phi(3\epsilon)+\Phi(\epsilon)}{2}$, and this inequality is implied by concavity of $\Phi$. Therefore, a cost-benefit ratio always exists that will permit this equilibrium, but the constraint is least restrictive for moderate levels of perception error.

We studied equation (83) numerically assuming $\Phi$ was a cumulative normal distribution and verified that it is positive but approaches zero at both high and low values of $\epsilon$. We also found that the maximum constraint is $\frac{r}{c} < \frac{1}{3}$ which occurs at approximately $\epsilon \approx 0.86$. These results match the equilibria shown in Fig.1 of the main text.

# 3 Numerical Extensions of the Model

The binomial error structure used in the analytical model of the previous sections was deliberately simple to make tractable closed-form characterizations of the conditional probabilities. Here we allow the distribution of the perception error to be either binomial or normal with mean 0 and standard deviation $\epsilon$. We assume that $a$ is distributed standard normal and we permit $k$ to take on any real value.

To produce numerical results from this model, we used the code shown in the next section, written in the R language. This code allows us to find equilibria via two methods in order to cross-check results and verify the analytical solutions.

In the "dynamic" method, we randomly match each individual in a population of constant size with another individual to play one iteration of the game (a "generation"). We then keep each individual with a probability of survival that is proportional to their game payoff. Individuals that are removed are replaced by surviving types with a probability that is proportional to their prevalence in the surviving population. At each generation we also randomly choose a fraction of the population to replace with a mutant type drawn from the set of all possible types. A dynamic plot (see Supplementary Fig.1) shows the distribution of surviving types in the population at fixed intervals over the course of a pre-specified number of generations.

In the "best reply" method, we find an iterated best reply equilibrium by starting with a population with a pre-specified set of types. The initial population can either be monomorphic or a mix of two types. We then identify the invading type that earns the highest expected payoff against the population. If this new type earns a lower payoff than the population average, then the population is at an equilibrium. Otherwise if the invading type beats all types in pairwise competition, then we replace the population with the invading type. If the invading type does not beat all types but nevertheless beats the population average then we replace the type that does more poorly with the invading type. We repeat this procedure until no invader can invade or until a previously-tested population is repeated.

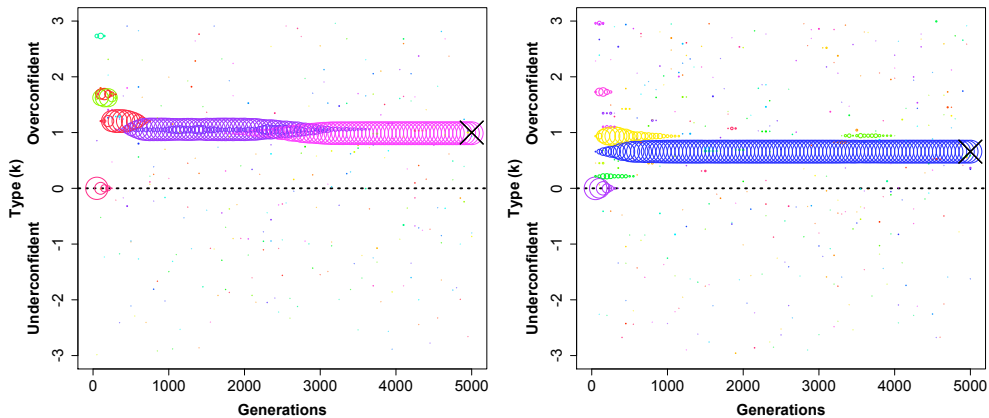

Supplementary Figure 1: Sample runs of the dynamic method for $r = 2$, $c = 1$, $\epsilon = 1$, where we start with a population of 1000 in which all individuals are unbiased ($k = 0$). The left panel shows results for the binomial model, and the right panel shows results for the normal model. The size of each circle is proportional to number of that type in the population at that generation. Colors are assigned randomly to each type. The X indicates the equilibrium predicted by analytic results (for the binomial model) and the best reply method (for both the binomial and normal models).

Both methods generate the same results, but we specifically use the best-reply method to construct the figures shown in the main text.

We systematically explored the parameter space defined by $0 < \epsilon < 5$, $0 < r < 5$, and $0 < c < 5$ for populations of size 1000, drawing 10,000 sets of values in these ranges from a uniform distribution. These values remained fixed for each simulation. The results of these simulations are shown in Fig.1 in the main text, and they indicate that both methods reproduce the analytical results when there is a binomial perception error, and that the equilibrium value $k^* > 0$ for all $\frac{r}{c} > 0.7$ when there is any positive normal perception error (regardless of its magnitude).

In general, both methods produce the same equilibria, subject to the number of generations used in the dynamic method and the accuracy of the integral of the normal distribution of $a$ used to calculate the expected value in the best reply method (see Supplementary Fig. 1 for an example). We manually checked discrepancies arising from parameter values near bound-ary thresholds, or the dynamic simulation not running long enough, by examining the set of invader payoffs for all types to determine which of the proposed equilibria were valid.

Under the assumption of a binomial perception error, the starting population sometimes determines the equilibrium that can be reached. For example, to find equilibria that contain at least one unbiased type ($k = 0$), we must start with a population at or near $k_j = 0$. All parameter combinations that yield an unbiased type as part of one equilibrium also yield a second equilibrium that contains no unbiased types (see Fig.1 of the main text). Specifically, when a monomorphic unbiased population is an equilibrium, so is a mixed population of underconfident ($k = -\epsilon$) and overconfident ($k = \epsilon$) individuals. And when a mix of unbiased and very underconfident ($k = -2\epsilon$) individuals is an equilibrium, so is a monorphic population of underconfident individuals ($k = -\epsilon$). To find these second equilibria, we typically start with a population of two extreme values (e.g. $k = \{-10, 10\}$).

## 3.1   R Code for the Numerical Models

```
## FUNCTION TO FIND SPECIFIC PAYOFFS WHEN EACH INDIVIDUAL
## PLAYS ONE OTHER INDIVIDUAL IN THE POPULATION

simkvec<-function(k,error) {

 # randomly match each i with a j
 pairs<-matrix(sample(k),ncol=2) # generate random pairs
 k_i<-c(pairs[,1],pairs[,2]) # create vector of types for individual i
 k_j<-c(pairs[,2],pairs[,1]) # create vector of types for individual j

 # errors and advantages
 a<-rnorm(n) # endowment advantage of i over j
 if(error=="normal") {
  e_i<-rnorm(n,0,e) # errors in i perception of j
  e_j<-rnorm(n,0,e) # errors in j perception of i
 } else {
  if(error=="binomial") {
   e_i<-sample(c(e,-e),n,replace=T) # errors in i perception of j
   e_j<-sample(c(e,-e),n,replace=T) # errors in j perception of i
  } else {
   return("need to specify binomial or normal error")
  }
 }

 # probabilities of claiming and winning
```

```
 i_claims <- ( a + e_i + k_i) > 0 # determine if i makes a claim
 j_claims <- (-a + e_j + k_j) > 0 # determine if j makes a claim
 i_wins <- a > 0 # determine if i beats j

 # payoffs
 # r   if i fights and j does not fight,
 # r-c if j fights and i wins, and
 #  -c if j fights and i loses,
 # 0   otherwise
 payoffs_i <- i_claims * ifelse(j_claims, i_wins*r - c, r)

 # return payoffs
 return(cbind(k_i, payoffs_i))
}

## FUNCTION TO APPROXIMATE E(PAYOFF) TO k_i AGAINST A SPECIFIC k_j

simkpair<-function(k,error) {

 # advantage distribution
 a <- (-5*nn):(5*nn)/nn # advantage to i
 pr_a <- dnorm(a) # probability density for a
 pr_a <- pr_a / sum(pr_a) # normalize density to create probabilities

 # probabilities of claiming and winning
 if(error=="normal") {
  i_claims <- pnorm( a + k[1], sd=e) # probability i makes a claim
  j_claims <- pnorm(-a + k[2], sd=e) # probability j makes a claim
 } else if(error=="binomial") {
  i_claims <- 0.5*( a + k[1] + e > 0) + 0.5*( a + k[1] - e > 0) # probability i makes a claim
  j_claims <- 0.5*(-a + k[2] + e > 0) + 0.5*(-a + k[2] - e > 0) # probability j makes a claim
 } else {
  return("need to specify binomial or normal error")
 }
 i_wins <- a > 0 # determine if i beats j
 i_ties <- which(a == 0) # determine when i ties j

 # payoffs
 # r   if i fights and j does not fight,
 # r-c if j fights and i wins, and
 #  -c if j fights and i loses,
 # 0   otherwise
 payoffs_i <- i_claims * (j_claims * (i_wins * r - c) + (1 - j_claims) * r)

 # if i and j tie, there is a 0.5 chance i wins
 payoffs_i[i_ties] <- i_claims[i_ties] *
   (j_claims[i_ties] * (0.5 * r - c) + (1 - j_claims[i_ties]) * r)

 # return payoffs
 return(sum(payoffs_i * pr_a ))
}

## FUNCTION TO GET PAYOFFS FOR AN INVADER AND 2 EXISTING TYPES IN A POPULATION

# get payoffs for an invader and 2 types in a population
```

```
simk3<-function(x,y,error) {

 # if type 1 and 2 are the same, then do not calculate mixture
 if(y[1] == y[2]) {
   p_pop<-simkpair(c(y[1],y[1]),error=error) # type 1 payoff vs. self
   mix<-1 # arbitrarily assign type 1 to be the winning type

 # otherwise, calculate mixture
 } else {

  # calculate payoffs when each population type plays self and other
  p11<-simkpair(c(y[1],y[1]),error=error) # type 1 payoff vs. type 1
  p12<-simkpair(c(y[1],y[2]),error=error) # type 1 payoff vs. type 2
  p21<-simkpair(c(y[2],y[1]),error=error) # type 2 payoff vs. type 1
  p22<-simkpair(c(y[2],y[2]),error=error) # type 2 payoff vs. type 2

  # calculate proportion of type 1 in equilibrium mixture
  if(p11>=p21&p12>=p22) {
   mix<-1
  } else {
   if(p21>p11&p22>p12) {
    mix<-0
   } else {
    mix<-(p22-p12)/(p11-p12-p21+p22)
   }
  }

  # payoff of population
  p_pop<-mix*p11+(1-mix)*p12
 }

 # payoff of each invader type vs. population
 p_inv<-rep(NA,length(x))
 for(i in 1:length(x)) {
  p_inv[i]<-    mix * simkpair(c(x[i],y[1]),error=error) +
(1-mix) * simkpair(c(x[i],y[2]),error=error)
 }

 # return mixture and invader and population payoffs
 return(cbind(mix,p_inv,p_pop))
}

## FUNCTION USED BY BEST REPLY FUNCTION
simk3optimize <- function(x,y,error=error) simk3(x,y,error=error)[2]

## FUNCTION TO IDENTIFY EQUILIBRIA BASED ON ITERATED BEST REPLY

bestreply<-function(khats,error) {

 khatslist<-paste(khats,collapse=" ") # initialize poulation list
 continue<-TRUE # initialize continuation variable

 # start loop to search for best replies
 while(continue) {
  print(khats)
```

```
# find maxima in various regions less than, greater than, and near 0
# we do this because the payoff curve is sometimes kinked for the binomial case
# at -2e, -e, 0, e, and 2e
o1 <-  optimize(simk3optimize, c(-5,-3*e/2), y=khats, maximum = T, error=error)
o2 <-  optimize(simk3optimize, c(-3*e/2,-e/2), y=khats, maximum = T, error=error)
o3 <-  optimize(simk3optimize, c(-e/2,e/2), y=khats, maximum = T, error=error)
o4 <-  optimize(simk3optimize, c(e/2,3*e/2), y=khats, maximum = T, error=error)
o5 <-  optimize(simk3optimize, c(3*e/2,5), y=khats, maximum = T, error=error)

objectives<-c(o1$objective,o2$objective,o3$objective,o4$objective,o5$objective)
maxima<-c(o1$maximum,o2$maximum,o3$maximum,o4$maximum,o5$maximum)

# choose invader with highest maximum
invader<-maxima[which.max(objectives)]

# evaluate best invader vs. population
s<-simk3(invader,y=khats,error=error)

# if invader does better than population....
if(s[2]>=s[3]) {
 # check current type against a mix of invader and other current type
 s1<-simk3(khats[1],y=c(khats[2],invader),error=error)
 s2<-simk3(khats[2],y=c(khats[1],invader),error=error)
 # if invader dominates both populations, create whole pop from invaders
 if(s1[1]==0&s2[1]==0) {
  khats<-c(invader,invader)
 } else {
  # replace type that does worst against invader
  if(s1[2]>=s2[2]) khats<-sort(c(khats[1],invader))
  if(s2[2]>s1[2]) khats<-sort(c(khats[2],invader))
 }
}

# add this population to the list (round to 3 digits)
khatslist<-c(khatslist,paste(round(khats,3),collapse=" "))

# if this population was already evaluated, stop searching
if(khatslist[length(khatslist)]%in%khatslist[1:(length(khatslist)-1)]) continue<-FALSE

}

# print approximate solution
return(khats)
}

## STEP 1: SET PARAMETERS FOR SIMULATION

n<-1000          # number of individuals in population
c<-1             # cost of fighting
r<-2             # reward to winning
e<-1             # size of error perception
d<-0.5           # approx. fraction of population that dies each time
g<-5000          # number of generations
m<-0.001         # mutation rate
gg<-50           # how many generations to skip between plot points
error<-"binomial" # choose binomial or normal perception error
start<-"unbiased" # choose starting population
```

```
# initial distribution of possible k thresholds in population
# NOTE: IF THERE ARE TWO EQUILIBRIA, STARTING VALUES WILL DETERMINE
# WHICH ONE HAPPENS.

if(start=="unbiased") {
 # start with unbiased population
 k<-rep(0,n)
} else {
 # otherwise start with extreme mixed population
 k<-c(rep(-10,round(n/2)),rep(10,n-round(n/2)))
}

# initialize plot
par(mai=c(1,1.2,0.1,0.1))
colors<-rainbow(6001)[sample(6001)] # randomly assign color to each type
plot(c(-1000,g*2),c(0,0),xlim=c(0,g),ylim=c(-3,3),type="l",lty="dotted",
     xlab="Generations",font.lab=2,cex.lab=1.5,cex.axis=1.5,lwd=3,
 ylab="Type (k)\nUnderconfident                Overconfident")

# loop through each generation
for(i in 1:g) {

 # get payoffs
 kp<-simkvec(k,error=error)
 k<-kp[,1]
 p<-kp[,2]

 # plot payoffs compared to k, color fixed for each k
 if(i%%gg==0) {
  tab<-table(k)
  points(rep(i,length(tab)),names(tab),pch=1,cex=5*sqrt(tab/n),
         col=colors[(as.numeric(names(tab))+3)*1000+1])
 }

 # normalize payoffs to create probability of remaining in population
 # proportional to payoff
 p<-(p-min(p))/(max(p)-min(p))
 p<-1-2*(1-p)*d

 # decide which individuals to remove from population
 dead<-which(p<runif(n,0,1))

 # replace dead strategies with live strategies
 if(length(dead)>0) k[dead]<-sample(k[-dead],length(dead),replace=T)

 # create mutant types, sampling from whole space
  mutant<-sample(n,round(n*m))
  k[mutant]<-round(runif(1,-3,3),3)

}

## STEP 2: ADD ITERATED BEST REPLY EQUILIBRIA
nn<-1000 # number of samples of normal distribution (higher = slower, more accurate)

# search near unbiased populations
b1<-bestreply(c(0,0),error=error)
```

```
# search again far from unbiased populations
b2<-bestreply(c(-10,10),error=error)

# print approximate equilibria (note, sometimes they will be different,
# but if they are nearby they are probably technically the same one)
print(b1)
print(b2)

# add best reply equilibrium points to plot
points(c(g,g),b1,pch=4,cex=5,col="black")
points(c(g,g),b2,pch=4,cex=5,col="black")
```

## 3.2   Interactions Between Three Individuals

One potential concern about the model is that it might generate results that are idiosyncratic to interactions between two individuals, important and common as these are in nature. Although it is beyond the scope of this article to generalize the results to $n$ individuals, we have adapted the numerical simulation to allow for interactions between three individuals.

As in the numerical model with two individuals, we set the distribution of the perception error to be normal with mean 0 and standard deviation $\epsilon$. We also assume that the advantage $a$ is distributed standard normal. Individuals only choose to claim a resource if they think they could win a conflict over it, and conflict only occurs if at least two of the three individuals claim the resource. All individuals in a conflict pay a cost $c$ and the winner of the conflict must beat all other claimants, which they must fight in sequence, one after the other. Furthermore, when one individual claims a resource and the other two do not, then the resource goes to the claimant. If none of the three individuals claims the resource, then all payoffs are 0.

We systematically explored the parameter space defined by $0 < \epsilon < 5$, $0 < r < 5$, and $0 < c < 5$ for populations of size 1000, drawing 10,000 sets of values in these ranges from a uniform distribution. Supplementary Fig. 2 reveals that the fundamental relationship between equilibrium confidence and the benefit-cost ratio remains intact. The main difference is that higher perception errors are required to reach extreme values of $k^*$ than in the two individual case. Otherwise, the threshold at which overconfidence becomes sustainable (approximately $\frac{r}{c} > 0.7$) appears to be similar, there are no

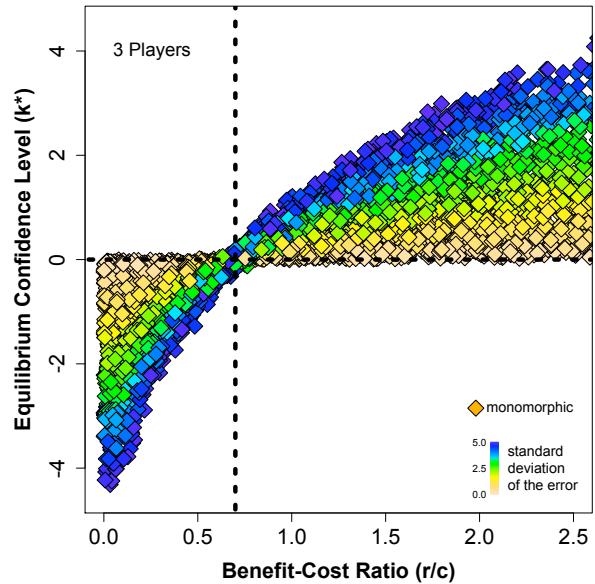

Supplementary Figure 2: Equilibrium levels of confidence ($k^*$) for varying benefit-cost ratios $\frac{r}{c}$ and varying degrees of uncertainty ($\epsilon$) about the capabilities of competitors in the model extension with three individuals competing for a resource. Notice the similarity to Figure 1 in the main text, where overconfidence is sustainable at approximately $\frac{r}{c} > 0.7$.

mixed equilibria, and the unbiased equilibrium is still rare. To produce numerical results from the model with three individuals, we altered the simkvec function as follows:

```
simkvec<-function(k) {

# randomly match each i, j, h
trios<-matrix(sample(k),ncol=3) # generate random trios
k_i<-trios[,1]
k_j<-trios[,2]
k_h<-trios[,3]

# errors and advantages
theta_i<-rnorm(n/3,0,sqrt(2)/2) # capability of i
theta_j<-rnorm(n/3,0,sqrt(2)/2) # capability of j
theta_h<-rnorm(n/3,0,sqrt(2)/2) # capability of h

a_ij<-theta_i-theta_h # endowment advantage of i over j
a_ih<-theta_i-theta_h # endowment advantage of i over h
```

```
a_jh<-theta_j-theta_h # endowment advantage of j over h

e_ij<-rnorm(n/3,0,e) # errors in i perception of j
e_ih<-rnorm(n/3,0,e) # errors in i perception of h
e_ji<-rnorm(n/3,0,e) # errors in j perception of i
e_jh<-rnorm(n/3,0,e) # errors in j perception of h
e_hi<-rnorm(n/3,0,e) # errors in h perception of i
e_hj<-rnorm(n/3,0,e) # errors in h perception of j

# determine if i makes a claim
i_claims <- (( a_ij + e_ij + k_i) > 0) & (( a_ih + e_ih + k_i) > 0)
# determine if i makes a claim
j_claims <- (( -a_ij + e_ji + k_j) > 0) & (( a_jh + e_jh + k_j) > 0)
# determine if i makes a claim
h_claims <- (( -a_ih + e_hi + k_h) > 0) & (( -a_jh + e_hj + k_j) > 0)

# probabilities of winning
i_wins <- (a_ij > 0) & (a_ih > 0) # determine if i beats j and h
j_wins <- (-a_ij > 0) & (a_jh > 0) # determine if j beats i and h
h_wins <- (-a_ih > 0) & (-a_jh > 0) # determine if h beats i and j

# payoffs
# r    if one individual claims and others do not claim,
# r-c  if two or more individuals fight and a individual wins, and
#  -c  if two or more individuals fight and a individual loses,
# 0    otherwise
payoffs_i <- i_claims * ifelse(j_claims|h_claims, i_wins*r - c, r)
payoffs_j <- j_claims * ifelse(i_claims|h_claims, j_wins*r - c, r)
payoffs_h <- h_claims * ifelse(i_claims|j_claims, h_wins*r - c, r)

# return payoffs
return(cbind(c(k_i,k_j,k_h),c(payoffs_i,payoffs_j,payoffs_h)))
}
```

## 3.3 Asymmetric Costs

Another concern about the model is that it might generate results that are idiosyncratic to the assumption that costs are the same for losers as for winners. In response, we have adapted the numerical simulation to allow the loser's costs to be higher than the winner's costs.

As in the numerical model with symmetric costs, we set the distribution of the perception error to be normal with mean 0 and standard deviation $\epsilon$. We also assume that the advantage $a$ is distributed standard normal. Individuals that lose a conflict incur a cost $c$ as usual, but winners incur a smaller cost $0 < c_{win} < c$.

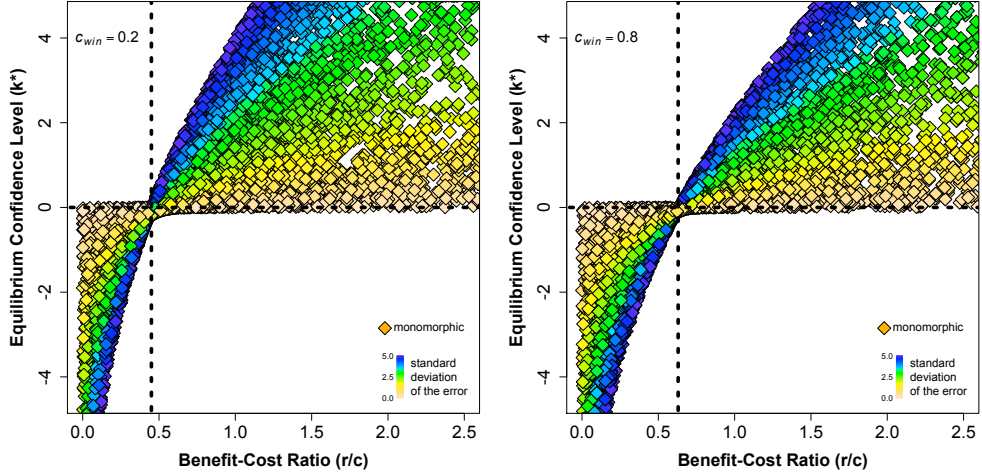

Supplementary Figure 3: Equilibrium levels of confidence ($k^*$) for varying benefit-cost ratios $\frac{r}{c}$ and varying degrees of uncertainty ($\epsilon$) about the capabilities of competitors in the model extension in which winners of conflicts incur a fraction of the cost of losers when competing for a resource. Notice the similarity to Figure 1 in the main text, except here overconfidence is sustainable for approximately $\frac{r}{c} > 0.45$ (where $c$ is the loser's cost) when $c_{win}/c = 0.2$ and approximately $\frac{r}{c} > 0.6$ when $c_{win}/c = 0.8$.

We systematically explored the parameter space defined by $0 < \epsilon < 5$, $0 < r < 5$, and $0 < c < 5$ for populations of size 1000, drawing 10,000 sets of values in these ranges from a uniform distribution. We evaluated two scenarios, one in which $c_{win}/c = 0.2$ and another in which $c_{win}/c = 0.8$, to determine whether asymmetric costs fundamentally altered the relationship between the benefit/cost ratio and equilibrium overconfidence.

Supplementary Fig. 3 shows that the fundamental relationship between equilibrium confidence and the benefit-cost ratio remains intact, there are no mixed equilibria, and the unbiased equilibrium is still rare. The main difference is that overconfidence becomes an equilibrium at a *lower* threshold as the cost to winners relative to losers ($c_{win}/c$) decreases. In other words, reducing costs incurred by winners makes it even more likely that overconfidence would evolve and be maintained.

To produce numerical results from the model with asymmetric costs, we altered the simkpair function as follows:

```
simkpair<-function(k,error) {

 # advantage distribution
 a <- (-5*nn):(5*nn)/nn # advantage to i
 pr_a <- dnorm(a) # probability density for a
 pr_a <- pr_a / sum(pr_a) # normalize density to create probabilities

 # probabilities of claiming and winning
 if(error=="normal") {
  i_claims <- pnorm( a + k[1], sd=e) # probability i makes a claim
  j_claims <- pnorm(-a + k[2], sd=e) # probability j makes a claim
 } else if(error=="binomial") {
  i_claims <- 0.5*( a + k[1] + e > 0) + 0.5*( a + k[1] - e > 0) # probability i makes a claim
  j_claims <- 0.5*(-a + k[2] + e > 0) + 0.5*(-a + k[2] - e > 0) # probability j makes a claim
 } else {
  return("need to specify binomial or normal error")
 }
 i_wins <- a > 0 # determine if i beats j
 i_ties <- which(a == 0) # determine when i ties j

 # payoffs
 # r    if i fights and j does not fight,
 # r-c  if j fights and i wins, and
 #  -c  if j fights and i loses,
 # 0    otherwise
 payoffs_i <- i_claims * (j_claims * ifelse(i_wins, r - c_win, - c) + (1 - j_claims) * r)

 # if i and j tie, there is a 0.5 chance i wins
 payoffs_i[i_ties] <- i_claims[i_ties] *
    j_claims[i_ties] * (0.5 * (r - c_win) + 0.5 * (- c) + (1 - j_claims[i_ties]) * r)

 # return payoffs
 return(sum(payoffs_i * pr_a ))
}
```

# 4 The Hawk-Dove Game is a Special Case

The well-known Hawk-Dove game is a special case of the model we present here. Under the assumption that only two behavioural types are possible, $k_i = \infty$ (Hawk) and $k_i = -\infty$ (Dove), the following table shows the expected fitness $E(f)$ in our model for each pairing of outcomes:

Supplementary Table 1: Expected fitness in the special case where $k \in \{\infty, -\infty\}$ as in the Hawk-Dove game

|  |  | Hawk $k_j = \infty$ | Dove $k_j = -\infty$ |
|---|---|---|---|
| Hawk | $k_i = \infty$ | $r/2 - c, r/2 - c$ | $r, 0$ |
| Dove | $k_i = -\infty$ | $0, r$ | $0, 0$ |

When $r > 2c$, this is not, strictly speaking, a Hawk-Dove game since a monomorphic population of Hawks ($k_j = \infty$) is the unique equilibrium if we only permit Doves as the alternative type. However, when $r < 2c$, a polymorphic population of Hawks ($k_j = \infty$) and Doves ($k_j = -\infty$) is possible.

Our model shows that overconfident individuals ($k_i = \epsilon$) can invade a population of Hawks. The expected fitness of the invading individual is

$$r/2 - \left(\frac{3}{4} - \Phi(2\epsilon)\right) c \tag{85}$$

which is clearly always greater than the fitness of a Hawk in a population full of Hawks (shown in the table above) since $\frac{3}{4} - \Phi(2\epsilon) < 1$.

Similarly, overconfident individuals ($k_i = \epsilon$) can invade a population of Doves. The expected fitness of the invading individual is

$$\left(\frac{1}{4} + \frac{1}{2}\Phi(2\epsilon)\right) r \tag{86}$$

which is clearly always greater than the fitness of a Dove in a population full of Doves, since $r > 0$.

Finally, a mixed equilibrium with a proportion of $p$ Hawks ($k_j = \infty$) and $1 - p$ Doves ($k_j = -\infty$) is always vulnerable to invasion by an overconfident individual ($k_i = \epsilon$). We can see this by noting first the mix at equilibrium is where $p(r/2 - c) + (1 - p)r = 0$ which simplifies to $p = 2r/(2c + r)$. Then we check the circumstances under which overconfident individuals beat both Hawks and Doves:

$$p \left( r/2 - \left( \frac{3}{4} - \Phi(2\epsilon) \right) c \right) + (1 - p) \left( \frac{1}{4} + \frac{1}{2} \Phi(2\epsilon) \right) r > 0 \qquad (87)$$

which, substituting for $p$, simplifies to:

$$\left( \frac{3}{4} - \frac{1}{2} \Phi(2\epsilon) \right) r + \left( \frac{3}{2} \Phi(2\epsilon) - \frac{1}{2} \right) c > 0 \qquad (88)$$

and this is always true since $r > 0$, $c > 0$, and $\Phi < 1$.

The intuition here is that Hawks always claim the resource, even in cases where the opponent's advantage is larger than the bias in an overconfident individual's assessment of the opponent. Under these conditions, overconfident individuals will avoid the cost of conflicts they are sure to lose and therefore will do better than Hawks. Similarly, Doves never claim the resource, even in cases where the opponent's advantage is smaller than the bias in an overconfident individual's assessment of the opponent. Under these conditions, overconfident individuals will gain the resource without cost and do better than Doves.

Interestingly, this means that when a technology for assessing opponents has evolved in a given population (even when assessment is biased!), the widely used Hawk-Dove game may not be the best model and may generate misleading predictions about outcomes. However, it is important to remember that there are many variations of the Hawk-Dove game that have been proposed, such as those that allow individuals to share resources with their opponent without conflict.

# 5 Levels of Conflict in Different Environments

A systematic exploration of the parameter space (as in Section 3 above) also shows a positive relationship between the benefit/cost ratio and the probability of conflict (Supplementary Fig. 4). It is not surprising that overconfidence is itself associated with greater levels of conflict—in fact that is close to our original assumption that overconfident individuals are more willing to claim resources they may not be able to win. However, we did not specify a direct relationship between resource benefits, costs, and conflict; the strong positive relationship suggests that environments with more valuable resources will generate more conflict.

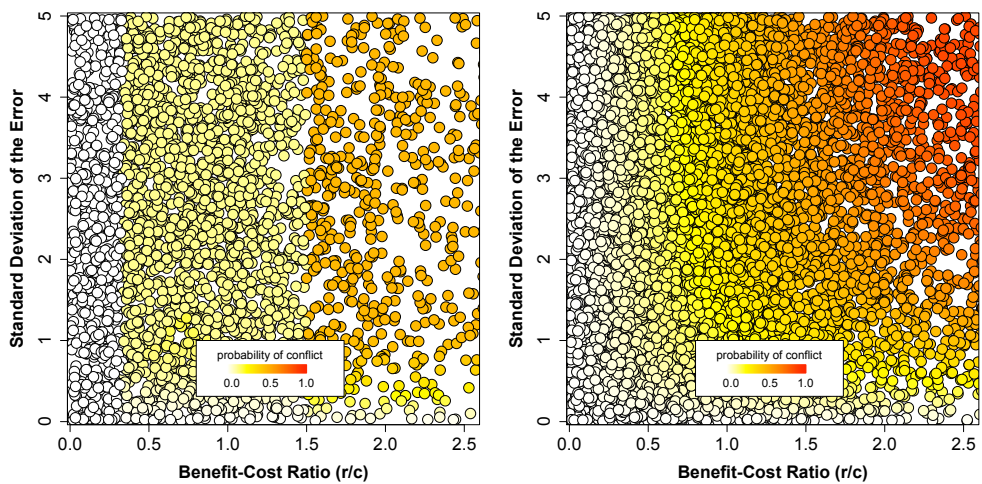

Supplementary Figure 4: Levels of conflict produced by equilibrium levels of confidence ($k^*$) and varying benefit/cost ratios ($r/c$) for different degrees of uncertainty ($\epsilon$) about the capabilities of competitors in the binomial (left panel) and normal (right panel) models. Each point shows the results from a single simulation where the cost, benefit, and degree of uncertainty were drawn from a uniform distribution. There were 10,000 simulations in total. The results show that conflict is increasing in both uncertainty and the relative value of contested resources ($r/c$).

The implication is that the environment $(r/c)$ in some sense dictates the optimal level of conflict that will maximize fitness over time and individual biases will adapt to match that level of conflict—even if that means sometimes initiating conflicts that they may lose.

# 6   A Note on Risk Preferences

In this section we show that individual behavior in the model is consistent with an alternative motivation for behavior that is based on risk preferences and the cognitively more demanding task of evaluating expected utility.

Suppose that individuals assume that all other individuals in the population are like themselves in that 1) they know the cost of conflict and the benefit of the resource, 2) they are capable of estimating the correct probability that the opponent will claim the resource, 3) they are capable of estimating the correct probability they would win a conflict, conditional on their error-prone subjective advantage assessment, and 4) they know the exact distribution of the perception error.

If all these conditions are met, then individuals can calculate the expected fitness of the risky choice $E(f_{risky})$ and compare this to a risk preference $\pi_i$ that defines how much additional fitness they require in expectation to justify taking such a risk. In our model, since the sure thing payoff of not making a claim is 0, then an individual with risk preferences makes a claim if $E(f_{claim}) > \pi_i$. This shows that risk tolerance decreases as $\pi_i$ increases.

To evaluate the expected value of the risky choice, recall that the variable $a$ denotes the advantage of individual $i$ over individual $j$. If $a > 0$ then $i$ would win a conflict between them. Neither individual observes $a$ because of perception error. Instead, $i$ perceives the advantage to be $s_i = a + \nu_i$ and $j$ perceives it to be $s_j = a + \nu_j$ (where $\nu$ is distributed $Pr(\nu = \epsilon) = 0.5$ and $Pr(\nu = -\epsilon) = 0.5$). Since the true underlying advantage $a$ is the same for both individuals, we can define individual $j$'s perception as a function of individual $i$'s perception. Moreover, the value depends on the distribution of the perception errors which yields four equally likely conditions:

$$
s_j = \begin{cases}
s_i & \text{if } \nu_i = \epsilon \text{ and } \nu_j = \epsilon \\
s_i + 2\epsilon & \text{if } \nu_i = \epsilon \text{ and } \nu_j = -\epsilon \\
s_i - 2\epsilon & \text{if } \nu_i = -\epsilon \text{ and } \nu_j = \epsilon \\
s_i & \text{if } \nu_i = -\epsilon \text{ and } \nu_j = -\epsilon
\end{cases}
\tag{89}
$$

Recall that in the overconfidence equilibrium of the binomial model, $k^* = \epsilon$ and therefore the decision to claim a resource if $s_i + k_i > 0$ implies the same decision is made if $s_i > -\epsilon$. Now we can calculate the expected fitness of claiming the resource conditional on different values of $s_i$.

If $s_i < -\epsilon$, then $s_j < \epsilon$ and $a < 0$. Under these conditions, individual $j$ always makes a claim and individual $i$ always loses a conflict. Since the cost of conflict $c > 0$ is a sure thing, there is no risk premium and $i$ never claims the resource regardless of risk preferences.

If $-\epsilon < s_i < 0$, then the probability that $j$ makes a claim is $Pr(s_j < \epsilon) = \frac{3}{4}$ since $s_j = s_i + 2\epsilon$ is the only one of the four conditions that yields a contradiction. Of the three conditions in which individual $j$ makes a claim, all three imply $a < 0$. Therefore, $j$ always wins a conflict. This means the fitness payoff to $i$ for making a claim is $\frac{1}{4}r - \frac{3}{4}c$. This is also the risk premium, since we are comparing this outcome to the sure-thing fitness of $0$ that results by not claiming the resource.

If $0 < s_i < \epsilon$, then the probability that $j$ makes a claim is $Pr(s_j < \epsilon) = \frac{3}{4}$ since $s_j = s_i + 2\epsilon$ is the only one of the four conditions that yields a contradiction. Of the three conditions in which $j$ makes a claim, two of them (when $s_j = s_i$) imply $a > 0$ and one of them ($s_j = s_i - 2\epsilon$) implies $a < 0$. Therefore, the probability that $j$ wins a conflict is $\frac{2}{3}$. This means the fitness payoff to $i$ for making a claim is $\frac{3}{4}(\frac{1}{3}(r-c)+\frac{2}{3}(-c))+\frac{1}{4}r$. Simplifying, the risk premium here is $\frac{1}{2}r - \frac{3}{4}c$.

If $\epsilon < s_i < 2\epsilon$, then the probability that $j$ makes a claim is $Pr(s_j < \epsilon) = \frac{1}{4}$ since $s_j = s_i - 2\epsilon$ is the only one of the four conditions that does not yield a contradiction. However, this condition ($s_j = s_i - 2\epsilon$) implies $a < 0$. Therefore, $j$ never wins a conflict. This means the fitness payoff to $i$ for making a claim is $\frac{1}{4}(r - c) + \frac{3}{4}r$. Simplifying, the risk premium is $r - \frac{1}{4}c$.

Finally, if $s_i > 2\epsilon$, then $s_j > \epsilon$. Under these conditions, individual $j$ never makes a claim which yields a sure thing payoff of $r$ when individual $i$ makes a claim. Since there is no uncertainty, there is no risk premium.

By inspection, it is easy to see that the maximum risk preference that would allow an individual to make the risky choice when $s_i > -\epsilon$ is $\pi_i < \frac{1}{4}r - \frac{3}{4}c$. Any individual with a risk preference $\pi_i > \frac{1}{4}r - \frac{3}{4}c$ would switch to the sure thing payoff in at least one case (when $-\epsilon < s_i < 0$). This means that in the overconfidence equilibrium, any individual with such preferences would be less fit and could not survive, just as individuals with $k_i < \epsilon$ could not survive. So in equilibrium only individuals with sufficiently low $\pi_i$ (i.e. those of sufficient risk tolerance) are sustainable.

What these calculations show is that the simple overconfidence heuristic can cause individuals to behave *as if* they are calculating the expected outcome of a risky choice under a specific set of assumptions about their opponents and comparing it to a required risk premium. As we discuss in the main paper, evolution may have favoured the simpler mechanism to achieve the same result.

# 7   Populations with a Mix of Unbiased Types ($k_j = 0$) and Very Overconfident Types ($k_j = 2\epsilon$) Are Not Stable

Overconfident individuals can invade a population if they get as much (or more) fitness when paired against a member of that population as that member gets when paired against itself. Suppose we have an underconfident population where each individual has confidence level $k_j = -\epsilon$. Under what conditions could an overconfident individual ($k_i = \epsilon$) invade this population? We can use equations (1) and (5) to derive an expected fitness for an overconfident individual that is matched with an underconfident individual:

$$\left(\frac{3}{8} + \frac{1}{4}\Phi(2\epsilon)\right)r - \left(\frac{1}{4}\Phi(2\epsilon) - \frac{1}{8}\right)c \tag{90}$$

This fitness is greater than the average fitness for individuals in a monomorphic population of underconfident individuals (equation (6)) when:

$$\frac{r}{c} > \frac{1}{3} \tag{91}$$

Now suppose that the population originally contains individuals that are all unbiased ($k_j = 0$). We use equations (1) and (5) to derive an expected fitness for an overconfident individual that is matched with an unbiased individual:

$$\left(\frac{3}{8} + \frac{1}{4}\Phi(\epsilon)\right)r - \left(\frac{1}{2}\Phi(2\epsilon) + \frac{1}{4}\Phi(\epsilon) - \frac{3}{8}\right)c \tag{92}$$

This fitness is greater than the average fitness for individuals in a monomorphic population of unbiased individuals (equation (6)) when:

$$\frac{r}{c} > \frac{\Phi(2\epsilon)}{\Phi(\epsilon) - \frac{1}{2}} - \frac{1}{2} \tag{93}$$

which is clearly always possible when $\epsilon$ is positive since $r$ and $c$ are positive. For very small $\epsilon$ the benefit/cost ratio needed is very large, but this quickly asymptotes to $\frac{3}{2}$ as $\epsilon$ increases (for example, the condition is $\frac{r}{c} > 2.37$ when $\epsilon = 1$ if $\phi$ is distributed standard normal). The implication is that a benefit/cost ratio always exists in which overconfident individuals can invade a population of unbiased individuals, and for moderate degrees of uncertainty about opponent capabilities the benefit/cost ratio need not be very high to ensure that overconfident individuals can do better than their unbiased peers.

Finally, we show that overconfident individuals $(k_j = \epsilon)$ can always invade a mixed population of very overconfident $(k_j = 2\epsilon)$ and unbiased $(k_j = 0)$ individuals. Given an equilibrium population with proportion $0 < p < 1$ unbiased types and proportion $1 - p$ very overconfident types, the following statement must be true in order for an overconfident type to invade:

$$pE(f|k_i = \epsilon, k_j = 0) + (1-p)E(f|k_i = \epsilon, k_j = 2\epsilon) >$$
$$pE(f|k_i = 2\epsilon, k_j = 0) + (1-p)E(f|k_i = 2\epsilon, k_j = 2\epsilon) \tag{94}$$

We use equations (1) and (5) to derive each of the payoffs in this equation, and simplifying yields the following inequality:

$$(\Phi(\epsilon) - \Phi(0))rp + (2(\Phi(3\epsilon) - \Phi(2\epsilon)) + (\Phi(\epsilon) - \Phi(0))))cp+$$
$$2((\Phi(3\epsilon) - \Phi(2\epsilon)) + (\Phi(\epsilon) - \Phi(0)))c(1-p) > 0 \tag{95}$$

Notice that all terms on the L.H.S. are positive, and therefore the statement is always true. Therefore no mixed population of very overconfident $(k_j = 2\epsilon)$ and unbiased $(k_j = 0)$ types can resist invasion by an overconfident individual $(k_j = \epsilon)$.